\documentclass[10pt]{iopart}
\usepackage{graphicx}
\usepackage{cite}
\usepackage{xcolor}
\usepackage{subcaption}
\usepackage{url}

\usepackage{enumerate}

\usepackage{mathtools}
\usepackage{amsmath}
\usepackage{amssymb}

\usepackage{float}

\begin{document}

\title[Flattening of Safety Factor Profile]{Turbulence-Induced Safety Factor Profile Flattening at Rational Surfaces in Tokamaks with Low Magnetic Shear}

\author{Arnas Vol\v{c}okas$^1$, Justin Ball$^1$, Giovanni Di Giannatale$^1$, Stephan Brunner$^1$}

\address{$^1$ Ecole Polytechnique F\'{e}d\'{e}rale de Lausanne (EPFL), Swiss Plasma Center (SPC), CH-1015 Lausanne, Switzerland}
\ead{Arnas.Volcokas@epfl.ch}
\vspace{10pt}
\begin{indented}
\item[]December 2024
\end{indented}

\begin{abstract}
In this paper, we investigate the effects of ion-scale turbulence-generated currents on the local safety factor profile under conditions of low magnetic shear and proximity to rational surfaces, relevant to Internal Transport Barrier (ITB) formation. 
Our results show that turbulent currents can generate stationary zonal magnetic potential corrugations, producing a stepped safety factor profile with extended regions of zero magnetic shear.
This change significantly affects turbulence self-interaction, resulting in a substantial decrease in turbulent transport, indicating a potential triggering mechanism for transport barrier formation.
\end{abstract}

%
\vspace{2pc}
%
%
%
%

\section{Introduction}

\phantom{.}

An outstanding problem in magnetic confinement fusion is the finite confinement time limited by turbulent transport, which obstructs fusion conditions and poses a major hurdle in the commercialization of fusion energy. 
Plasma core performance, however, can be greatly enhanced by the formation of Internal Transport Barriers (ITBs) \cite{Wolf2003, Connor2004, Ida2018}, which are characterized by sharp and localized steepening of the plasma pressure profile near the core. 
At the same time, core temperature increases, and turbulent transport is reduced, enabling more favorable fusion conditions.
However, despite extensive experimental studies and well-documented observation of formation conditions across multiple devices \cite{Wolf2003, Connor2004, Ida2018, Garbet2010, Joffrin2002a, Tala2006}, the fundamental theoretical understanding of ITB formation remains limited. 
This knowledge gap impedes the potential utilization of ITBs in future fusion devices. 

From empirical observations, it is known that the safety factor profile plays a critical role in ITB triggering.
ITBs are preferentially formed in plasmas with reversed magnetic shear $q$ profiles, i.e. profiles characterized by negative magnetic shear in the inner radial region of the plasma and positive shear in the outer region, and with the profile minimum $q_{min}$ near a rational surface \cite{Lao1996, Eriksson2002, Joffrin2002, Joffrin2002a, Austin2006, Waltz2006 ,Staebler2018}. 
This offers an appealing parameter regime for numerical investigation, where the topology of the magnetic field plays a critical role. 
This work aims to numerically study the effects of turbulence-generated currents and associated magnetic field contributions to the background safety factor profile in this parameter regime, and their role in reducing turbulent transport and triggering ITBs.

In fusion plasmas, well-known mechanisms exist by which turbulence can generate parallel currents, either through the parallel nonlinearity (usually a small effect) or the radial flux of parallel electron momentum due to $E \times B$ drift and magnetic flutter \cite{Hinton2004}. 
The turbulent generated \textit{stationary} currents are most prominent around low-order rational surfaces where the radial variation in turbulence intensity leads to a residual contribution to parallel Reynolds stress \cite{Gurcan2010}.
This, in turn, contributes to the parallel electron momentum flux, driving the parallel current.
In the past, turbulent current generation has been explored in numerical simulations, primarily  using global codes \cite{Miyato2007, Wang2019, Chen2021}.
More recently their existence under electron temperature gradient (ETG) turbulence conditions has been confirmed experimentally \cite{Li2022}. 
However, little attention was given to a feedback mechanism by which turbulent generated currents can locally modify the safety factor profile and lead to changes in turbulence itself.

Our interest is particularly focused on the effects of turbulent currents in the low magnetic shear regime as turbulence-generated currents can more easily modify (and in particular flatten) the background safety factor profile when magnetic shear is weak. 
It is important to note that, in the gyrokinetic ordering considered in this work, plasma turbulence can alter the gradients of plasma profiles by creating stationary corrugations.
The gradients of these corrugations can compete with the existing background profile gradients, while the actual values of the plasma background profiles remain virtually unchanged.
As a result, turbulence can more effectively compete with background gradients when they are low.
Additionally, recent numerical work has demonstrated that when magnetic shear is weak or zero, turbulent eddies can extend along the magnetic field lines for many poloidal turns (when kinetic electron physics is considered) \cite{Volcokas2023}.
This, combined with the recognized role of parallel self-interaction in regulating turbulence \cite{Ball2019, Ajay2020, Volcokas2023, Volcokas2024_PartOne_L, Volcokas2024_PartTwo_NL}, suggests that the self-interaction of ultra long turbulent eddies at rational surfaces can create radial turbulence intensity variation and lead to a parallel current drive in the vicinity of these surfaces.

In this work, we study the effects of stationary parallel current profiles originating from ion temperature gradient (ITG) turbulence near rational surfaces with low magnetic shear. 
In particular, we demonstrate that these self-generated currents cause significant changes in the local safety factor profile by creating stationary zonal magnetic potential corrugations. 
This results in the flattening of the local safety factor profile around rational surfaces and the creation of a radial region with zero magnetic shear.
Furthermore, these safety factor profile corrugations lead to reduced transport, presenting an additional mechanism that may contribute to ITB formation. 
For our study, we use the flux tube version of the gyrokinetic GENE code \cite{Jenko2000, Gorler2011} with a standard linear safety factor profile for preliminary investigations and a non-uniform safety factor profile \cite{Ball2022} suitable for examining reversed shear configurations. 
We also cross-reference our GENE results with global gyrokinetic simulations obtained with the ORB5 code \cite{Lanti2020_ORB5}, which reveal very similar behavior around rational surfaces.
A more detailed analysis of these global results is presented in a companion paper \cite{DiGiannatale2024_GlobalReversedShear}.
Our findings suggest that local turbulence self-organization around rational surfaces with low magnetic shear, driven by strong parallel self-interaction, could be a key mechanism in ITB triggering.

The remainder of this paper is organized as follows.
In Sec. \ref{sec:theoretical_background}, we give a brief overview of the key theoretical concepts used throughout the work, namely the considered gyrokinetic model and in particular the flux tube simulation domain including its generalization to account for possible non-uniform magnetic shear \cite{Ball2022}, the turbulent parallel current drive, and the associated modifications of the safety factor profile.
The next four subsections are dedicated to presenting numerical results.
In Sec. \ref{sec:corrug_linear_profile}, we use standard low magnetic shear simulations with a linear safety factor profile and demonstrate the formation of a stepped safety factor profile due to a steady zonal component of $A_{\parallel}$. 
In Sec. \ref{sec:flat_small_perturb}, we use the novel non-uniform shear extension to the flux tube version of the GENE code to show that turbulent currents can completely cancel out small amplitude externally imposed safety factor profile perturbations. 
Following this, in Sec \ref{sec:q_profile_scan} we discuss a reversed safety factor profile scan across a low-order rational surface. 
In Sec. \ref{sec:global_sims}, we present \textit{pseudo-global} reversed shear simulations and compare our numerical results with global standard ORB5 simulations.
Lastly, we finish with an overview of this work and conclusions. 

\section{Theoretical background}\label{sec:theoretical_background}

\subsection{Gyrokinetic model}

In this work, we model plasma turbulence using gyrokinetics, a kinetic plasma model that self-consistently describes plasma evolution in the presence of strong magnetic fields. 
The basic idea is to apply the gyrokinetic ordering to the Vlasov equation to ``average-out" particle gyromotion.
A more detailed review of gyrokinetic theory can be found in Refs. \cite{Brizard2007, Abel2013}. 
For our calculations we use the standard version of the gyrokinetic equations where we consider fluctuations that are much smaller than the equilibrium values, thereby separating the full particle distribution function $\Bar{F}_{s}=\Bar{f}_{0,s}+ \Bar{f}_{1,s}$ into an equilibrium $\Bar{f}_{0,s}$ and a fluctuating $\Bar{f}_{1,s}$ part ($|\Bar{f}_{1,s}|/ \Bar{f}_{0,s} \ll 1$). 
The equilibrium distribution function is taken to be a local Maxwellian, assuming that the system is close to thermal equilibrium. 
The particle distribution functions are expressed in the gyrocenter variables $(\mathbf{x}, v_{\parallel}, \mu)$, where $\mathbf{x}$ is the gyrocenter position, $v_{\parallel}$ is the parallel component of the velocity and $\mu$ is the magnetic moment.
Additionally, we assume no parallel magnetic field fluctuations $\delta B_{\parallel} =0$ while maintaining the effects of perpendicular magnetic field fluctuations $\delta \mathbf{B}_{\perp}$ through the magnetic vector potential component $A_{1 \parallel}$ (here $\parallel$ and $\perp$ respectively refer to the directions parallel and perpendicular to the equilibrium magnetic field $\mathbf{B}_{0}$).
The strength of the electromagnetic effects is determined by plasma $\beta$, which is the ratio of magnetic pressure over thermal pressure.
With these assumptions and following Refs.  \cite{Lapillonne2010Thesis, Gorler2011}, the gyrokinetic equation can be written as
\begin{align}
    \label{eq:simple_GK_eq}
    & \frac{\partial}{\partial t} \left(\Bar{f}_{1,s} - \frac{Z_{s}e}{m_{s}}\Bar{A}_{1\parallel}\frac{\partial \Bar{f}_{0,s}}{\partial v_{\parallel}} \right) +\frac{B_{0}}{B_{0}^{*}}\mathbf{v}_{\Bar{\xi}} \cdot \left(\nabla \Bar{f}_{0,s}-\frac{\mu}{m_{s}v_{\parallel}} \nabla B_{0} \frac{\partial \Bar{f}_{0,s}}{\partial v_{\parallel}}\right) \notag \\
    & + \left[ v_{\parallel} \hat{\mathbf{b}} + \frac{B_{0}}{B_{0}^{*}}( \mathbf{v}_{\nabla B}+\mathbf{v}_{c}+\mathbf{v}_{\Bar{\xi}}) \right] \cdot \left(\nabla \Bar{f}_{1,s} - \frac{Z_{s}e}{m_{s} v_{\parallel}} \nabla \Bar{\phi}_{1} \frac{\partial \Bar{f}_{0,s}}{\partial v_{\parallel}} \right) \notag \\
    & - \frac{\mu}{m_{s}} \left(\hat{\mathbf{b}}+\frac{B_{0}}{v_{\parallel} B_{0}^{*}}\mathbf{v}_{c} \right) \cdot \nabla B_{0} \frac{\partial \Bar{f}_{1,s}}{\partial v_{\parallel}} =0,
\end{align}
where $B_{0}$ is magnitude of the equilibrium magnetic field $\mathbf{B}_{0}$, $\hat{\mathbf{b}} = (\mathbf{B}_{0}/B_{0})$ is the unit vector along the magnetic field, $B_{0}^{*}=\mathbf{B}_{0}^{*}\cdot \hat{\mathbf{b}}$ is the parallel component of $\mathbf{B}_{0}^{*} = \mathbf{B}_{0} + \nabla \times (\mathbf{B}_{0} v_{\parallel}/\Omega)$, $\mathbf{v}_{\nabla B} = [\mu / (m_{s} \Omega_{s} B_{0})] B_{0} \times \nabla B_{0}$ is the grad-B magnetic drift velocity, $\mathbf{v}_{c} = (v_{\parallel}^{2}/\Omega_{s}) (\nabla \times \hat{\mathbf{b}})_{\perp}$ is the curvature drift velocity, $\mathbf{v}_{\Bar{\xi}} = - (\nabla \Bar{\xi}_{1} \times \mathbf{B}_{0})/B_{0}^{2}$ is the generalized $\mathbf{E} \times \mathbf{B}$ drift (includes magnetic flutter if $A_{1,\parallel} \neq 0$), $m_{s}$ is the mass of species $s$ (ions and electrons), $e$ is the proton charge and $Z_{s}$ is the charge number of species $s$. Finally, $\Bar{\xi}_{1}$ is the gyroaveraged modified potential $\Bar{\xi}_{1} = \Bar{\phi}_{1}-v_{\parallel}\Bar{A}_{1 \parallel}$, where $\Bar{\phi}_{1}$ is the gyroaveraged electrostatic potential and $\Bar{A}_{1 \parallel}$ is the gyroaveraged parallel magnetic vector potential. 
Only the $\mathbf{v}_{\Bar{\xi}}$ nonlinearity is kept in Eq. \eqref{eq:simple_GK_eq}.
Quantities with an overbar are gyroangle averaged at a fixed guiding center position in a perturbed system. 

Instead of coupling Eq. \eqref{eq:simple_GK_eq} to the full Maxwell's equations, the system is closed by imposing quasineutrality
\begin{equation}
    \label{eq:quaineutrality}
    \sum_{s} Z_{s} e n_{1,s} \approx 0,
\end{equation}
where $n_{1,s}$ is the particle density of species $s$ calculated using the zeroth order velocity moment of the gyrocenter distribution function $\Bar{f}_{1,s}$.
The electrostatic potential $\phi_{1}$ is derived from the polarization term in the quasineutrality equation \cite{Lapillonne2010Thesis, Gorler2011}.

Finally, the magnetic vector potential component $A_{1, \parallel}$ is obtained from the parallel component of Amp\`{e}re's law
\begin{equation}
\label{eq:Amperes_law}
    - \nabla_{\perp}^{2} A_{1, \parallel} = \mu_{0} \sum_{s} j_{1 \parallel,s},
\end{equation}
where $j_{1 \parallel,s}$ is the parallel current carried by species $s$ and $\mu_{0}$ is the magnetic permittivity. 
The parallel current is defined as
\begin{align}
    \label{eq:parallel_current}
    j_{1 \parallel,s} & =  \frac{2\pi Z_{s}e}{m_{s}} \int B_{0}^{*} v_{\parallel} \Bar{f}_{1,s}(\mathbf{x}, v_{\parallel}, \mu) dv_{\parallel} d \mu \notag \\
    & - \frac{Z_{s}en_{0,s}\mu_{0}j_{0,\parallel}}{B^{2}_{0}} \left(\phi_{1}- \frac{B_{0}}{T_{0,s}} \int \Bar{\Bar{\phi}}_{1} \exp{\left(-\frac{\mu B_{0}}{T_{0,s}} \right)} d\mu \right) ,
\end{align}
where $j_{0,\parallel} = (\nabla \times \mathbf{B}_{0}/\mu_{0})\cdot \hat{\mathbf{b}}$ is the local MHD equilibrium parallel current and double overbar represents a double gyro-averaging.
In the following sections, we will drop subscript one from the potentials $\phi_{1} \rightarrow \phi$, $A_{1, \parallel} \rightarrow A_{\parallel}$ and parallel current $j_{1 \parallel,s} \rightarrow j_{\parallel, s}$ for simplicity.

\subsection{Flux tube simulation domain}

For the GENE simulations, we solve the gyrokinetic model in a narrow computational domain, that follows a set of neighboring background magnetic field lines, called a flux tube. 
In this section, we will briefly review the flux tube domain and field-aligned coordinate system as it is important for the following study. 
An in-depth discussion of the flux tube is given in the original paper \cite{Beer1995} and other works extending the model \cite{Ball2020, Ball2021, Ball2022, Volcokas2024_PartOne_L}.

The standard flux tube model considers a nonorthogonal, curvilinear, field-aligned Clebsch-type coordinate system defined in an axisymmetric tokamak as follows:
\begin{equation}
\label{eq:coordinates_flux_tube}
    x=x(\psi), \quad y=C_{y}\left[q(\psi)\chi-\zeta \right], \quad   z=\chi,
\end{equation}
where $x,y,z$ are referred to as the radial, binormal, and parallel coordinates, and $\psi, \chi, \zeta$ are the poloidal flux, straight field line poloidal angle, and toroidal angle respectively. 
$C_{y}$ is a normalization constant ensuring that $y$ has dimensions of length.
These coordinates are standard in gyrokinetic simulations, including the GENE code used here.\par

Due to turbulence anisotropy, the flux tube computational domain $\left( x,y,z \right) \in \left[ 0, L_{x} \right] \times \left[ 0, L_{y} \right] \times \left[ - \pi N_{pol}, \pi N_{pol} \right]$ is narrow in the perpendicular directions $x$ and $y$ compared to the full plasma cross-section but extended along the magnetic field lines. Here $L_{x}$ and $L_{y}$ are the widths of the radial and binormal directions respectively, and $N_{pol}$ quantifies the parallel length by the number of poloidal turns around the torus. 
In the local approximation based on the assumption of scale separation, equilibrium profiles like density, temperature, and safety factor are Taylor expanded to 1$^{st}$ order around the reference magnetic surface, meaning the value and the gradient of these quantities are constant across the flux tube and vary only with $z$. 

In the radial and binormal directions, periodic boundary conditions are
\begin{equation}
\label{eq:radial_boundary}
    A(x+L_{x}, y, z)= A(x,y,z),
\end{equation}
\begin{equation}
\label{eq:binormal_boundary}
    A(x, y+L_{y}, z) = A(x,y,z),
\end{equation}
where $A(x,y,z)$ represents any of the fluctuating quantities in gyrokinetics.
This allows us to represent such fluctuations with a Fourier series, whose Fourier coefficients are denoted $\hat{A}_{k_{x}, k_{y}}(z)$, where $k_{x}=k (2\pi/L_{x})$, $k \in \mathbb{Z}$, and $k_{y}=l (2 \pi / L_{y})$, $l \in \mathbb{Z}$, are the radial and binormal wavenumbers respectively.

Due to finite background magnetic shear $\hat{s}_{0}$, the so-called ``twist-and-shift" \cite{Beer1995} boundary condition 
\begin{equation}
\label{eq:parallel_BC}
    A(x, y + C_{y} q_{tot}(x) 2\pi N_{pol}, z+2 \pi N_{pol}) = A(x, y, z).
\end{equation}
is applied in the parallel direction, where $q_{tot}(x)$ is the total safety factor profile (which we will define later).
There are two important consequences of the twist-and-shift parallel boundary condition: (1) domain quantization and (2) (pseudo-)rational surfaces. 

First, the flux tube domain quantization condition reads \cite{Ball2020}
\begin{equation}
\label{eq:domain_quantization}
    L_{x} = \frac{M}{2 \pi N_{pol}|\hat{s}|} L_{y}=\frac{M}{k_{y,min}N_{pol}|\hat{s}|},
\end{equation}
where $M \in \mathbb{N}^{*}$ is a strictly positive integer.
However, the domain quantization condition does not hold when $\hat{s}=0$ case, for which one is free to choose $L_{x}$ and $L_{y}$ independently.

Second, at various specific radial locations, the magnetic field lines pass through the parallel boundary a certain finite integer number of times before connecting back onto themselves.
This corresponds to different rational surfaces, which can have different orders depending on how many poloidal turns they make before closing on themselves \cite{Ball2020}. 
The simulation domain always includes $M$ lowest-order rational surfaces (of order $N_{pol}$). 
This also means that there are $nM$ rational surfaces of order $nN_{pol}$, radially separated by distance $\Delta x = L_{x} /(nM)$.
The presence of these surfaces results in a unique turbulence behavior at low magnetic shear, due to turbulent self-interaction in the parallel direction that modifies turbulent transport \cite{Volcokas2024_PartTwo_NL}.
Note that normally the flux tube is constrained to cover only part of the full magnetic surface.
In this case, rational surfaces in the flux tube do not correspond to physical ones or have incorrect order.
Thus we call them \textit{pseudo}-rational surfaces.

On the other hand, in simulations with zero magnetic shear, the whole simulation domain is made up of identical magnetic surfaces. 
The effective order of the magnetic surfaces can be adjusted with the phase factor
\begin{equation}
\label{eq:phase_factor_C}
     C = e^{-i 2 \pi j \eta}.
 \end{equation}
This appears in the parallel boundary condition
\begin{equation}
    \hat{A}_{k_{x}, k_{y}} (z + 2 \pi N_{pol}) = \hat{A}_{k_{x}, k_{y}} (z) C,
\end{equation}
which can be derived from \eqref{eq:parallel_BC} by Fourier analysis and using $k_{y}= 2 \pi j / L_{y}, j \in \mathbb{Z}$.
Here $\eta \in (-0.5, 0.5]$ controls the binormal shift of the field lines at the parallel boundary as a fraction of the box width, while $\Delta y_{0} = \eta L_{y}$ corresponds to the real-space shift of the field lines.
Choosing $\eta =0, 1/2, ...$ corresponds to 1$^{st}$ order, 2$^{nd}$ order, etc. mode rational surfaces when $N_{pol}=1$.
The parallel boundary phase factor has been previously investigated at low and zero magnetic shear in Refs. \cite{Volcokas2023, St-Onge2023}, and an in-depth study is presented in Refs. \cite{Volcokas2024_PartOne_L, Volcokas2024_PartTwo_NL}.

In the later parts of this paper, we will use the binormal phase factor to adjust the radial locations of (pseudo-)rational surfaces, which will allow us to perform a scan in the magnetic field topology and safety factor values.
Note that physically this phase factor is determined by $q$.
The relationship between a binormal real-space shift $\Delta y_{0}$ of the equilibrium magnetic field lines and the corresponding offset $\Delta q$ from a rational value can be expressed as \cite{Volcokas2024_PartOne_L}
\begin{equation}
\label{eq:delta_q_notnormalized}
    \Delta q = \frac{\Delta y_{0}}{2 \pi C_{y} N_{pol}},
\end{equation}
where the offset is very small compared to the background safety factor values, i.e., $\Delta q / q_{0} \sim \rho^{*} \ll 1$.

\subsection{Non-uniform magnetic shear}\label{sec:non-uni_shear}

As ITB formation is facilitated by reversed shear configurations in experiments, we are interested in understanding the role of safety factor profile curvature in turbulence self-interaction. 
To investigate such configurations, we will go beyond the linearized safety factor profile usually considered in the standard flux tube model when imposing the parallel boundary condition in Eq. \eqref{eq:parallel_BC} and instead use a \textit{pseudo-global} formalism that retains higher-order terms in the Taylor expansion of the safety factor profile. 

Reference \cite{Ball2022} considered a hypothetical ECCD current source that could drive very narrow current layers. 
This source was used to self-consistently introduce an $L_{x}$-periodic modulation of the magnetic shear in gyrokinetic flux tube simulations, which allows the inclusion of local safety factor curvature effects.
Here we will briefly present the main features of the non-uniform shear formalism important for this work and refer readers interested in the details to reference \cite{Ball2022}.

The total safety factor profile $q_{tot}$ can be expressed as
\begin{equation}
\label{eq:total_imposed_q}
    q_{tot} = q_{0}+ \frac{q_{0}}{r_{0}} \hat{s}_{0} x + \Tilde{q}(x)+ \Delta q,
\end{equation}
where $\hat{s}_{0} = (r_{0}/q_{0}) dq/dx$ is the standard background magnetic shear that is constant across the flux tube, $r_{0}$ is the flux surface label at the center of the domain and $\Delta q$ is a possible small offset as given by Eq. \eqref{eq:delta_q_notnormalized}. 
Here $\Tilde{q}(x)$ is the non-uniformity containing all higher order safety factor profile terms. 
In the non-uniform magnetic shear formalism, $\Tilde{q}(x)$ is expressed as a Fourier series 
\begin{equation}
\label{eq:non_uniform_q}    \Tilde{q}(x) = \sum^{\infty}_{n=1} \frac{q_{0}}{r_{0}} \frac{L_{x}}{2\pi n} \left[ -  \Tilde{s}^{S}_{n} \cos{\left(\frac{2 \pi n}{L_{x}}x\right)} + \Tilde{s}^{C}_{n} \sin{\left(\frac{2 \pi n}{L_{x}}x \right)} \right],
\end{equation}
where $\Tilde{s}_{n}^{S}$ and $\Tilde{s}_{n}^{C}$ are sine and cosine Fourier coefficients of the magnetic shear modulation profile. 
This ensures that the imposed non-uniformity averages to zero across the radial domain to preserve the periodic boundary conditions. 

In this paper, we will use a non-uniform safety factor profile to
\begin{itemize}
    \item[] \textbf{(i)} show how the gyrokinetic system reacts to small non-uniformity in the safety factor profile,
    \item[] \textbf{(ii)} compare simulations with imposed stepped safety factor profiles against stepped safety factor profiles generated by turbulence in standard gyrokinetic simulations, and
    \item[] \textbf{(iii)} simulate turbulent transport in reversed shear configuration.
\end{itemize}

\subsection{Turbulent current drive}\label{sec:turbulent_current_drive}

From Amp\`{e}re's law in Eq. \eqref{eq:Amperes_law}, it is evident that parallel currents lead to the generation of the parallel component of the magnetic vector potential $A_{\parallel}$.
In a similar fashion to an externally imposed current source discussed in Sec. \ref{sec:non-uni_shear}, turbulent self-generated currents can lead to changes in the safety factor profile.
We will find that the main drive of the parallel current in our simulations is through the parallel electron momentum flux $\Pi_{e,\parallel}$, in agreement with Refs. \cite{McDevitt2017, Wang2019}.
Moreover, since there is no externally imposed mean parallel electron flow $U_{e,\parallel} = 0$, the only contribution to the parallel electron momentum flux comes from the electron parallel residual stress $\pi_{e, \parallel}$, i.e. $\Pi_{e,\parallel} \propto \pi_{e, \parallel}$.
We consider low magnetic shear conditions that have not been addressed in the past, where a large contribution to the electron parallel residual stress comes from the strong parallel self-interaction.

For this preliminary analysis of turbulent current drive, we will consider electrostatic turbulence and set $\beta \rightarrow 0$ such that electromagnetic effects can be neglected, i.e. we set magnetic potential to zero $\Bar{A}_{1 \parallel}=0$ in the gyrokinetic equation.
This allows us to study the drive of the current without its back-reaction on the magnetic equilibrium.
Additionally, we will only consider current generated by electrons since ion current drive is subdominant.
Since we consider TEM/ITG dominated turbulence, we neglect finite Larmour radius effects associated with electron gyroradius, which is much smaller than the turbulent spatial scale, and use drift kinetic electron approximation.
A detailed outline of the derivation and comparison to simulation results are presented in \ref{app:turb_current_drive}.

The final expression of the zonal parallel current evolution equation in the field-aligned flux tube coordinate system from \ref{app:turb_current_drive} is 
\begin{align}
\label{eq:current_eq_full}
     \partial_{t} \langle j_{1, \parallel} \rangle_{y,z} = & \frac{1}{C_{xy}}\partial_{x} \langle  \partial_{y} \phi_{1} j_{1, \parallel} \rangle_{y,z} \notag \\
     & - \partial_{x} \left\langle \frac{\mathcal{K}_{x}}{2\pi\Omega_{e}} C_{\mu_{0}}  (q_{1, \perp,e} + 2 q_{1,\parallel,e} +4 p_{0} u_{1,\parallel,e}) \right\rangle_{y,z} \notag \\
     & - \left\langle \frac{C_{\mu_{0}} C_{xy}}{J_{xyz} B_{0}} \frac{m_{e}}{4 \pi} \partial_{z}\left( T_{1,\parallel, e} n_{0,e} + T_{0,e}n_{1,e} \right) \right\rangle_{y,z} \notag \\
     & + \left\langle \frac{C_{\mu_{0}} C_{xy}}{J_{xyz} B_{0}}  \frac{q_{0} \partial_{z} \phi_{1}}{2 \pi}  n_{0e}  \right\rangle_{y,z} \notag \\
     &  + \left\langle \frac{C_{\mu_{0}} C_{xy} }{J_{xyz} B_{0}} \frac{\partial_{z} B_{0}}{B_{0}} \frac{m_{e}}{4 \pi} \left( T_{1,\parallel, e} n_{0,e} + T_{0,e}n_{1,e} \right) \right\rangle_{y,z} \notag \\
     & - \left\langle \frac{C_{\mu_{0}} C_{xy} }{J_{xyz} B_{0}} \frac{\partial_{z} B_{0}}{B_{0}} \frac{m_{e}}{4 \pi}   (T_{1, \perp,e } n_{0,e} + T_{0,e} n_{1,e})  \right\rangle_{y,z},
\end{align}
where $C_{xy}$ is a normalization factor, $\mathcal{K}_{x}$ is a geometric factor, $\Omega_{e}$ is the electron gyrofrequency, $C_{\mu_{0}}$ is a normalization constant, $q_{1,\perp,e}$ and $q_{1, \parallel, e}$ are perpendicular and parallel electron heat current densities respectively, $T_{1,\perp,e}$ and $T_{1, \parallel, e}$ are perpendicular and parallel electron temperatures respectively, $p_{0}$ is the background pressure, $n_{0,e}$ is the background electron density and $T_{0,e}$ is the background electron temperature.

Since we are only interested in the radial parallel current profile, a flux surface average was performed on the full space current evolution equation according to
\begin{equation}
\label{eq:flux_surface_avg}
    \langle ... \rangle_{y,z} \equiv \langle \langle ... \rangle_{y} \rangle_{z} = \frac{\int \int ... J_{xyz}(z) dy dz}{\int \int J_{xyz}(z) dy dz},
\end{equation}
where $\langle ... \rangle_{y}$ is the binormal average, $\langle ... \rangle_{z}$ is the average along the magnetic field lines and $J_{xyz}$ is the real space Jacobian.

In Eq. \eqref{eq:current_eq_full}, the first term on the right-hand side (RHS) is the radial divergence of parallel electron momentum flux, which we find to be the main contributor to the parallel electron current drive. 
This current drive mechanism requires radial inhomogeneity in turbulence \cite{Parra2015a}, a condition that can be met near rational surfaces. 
In contrast, the other terms in the equation, specifically those arising from the curvature, parallel acceleration, and mirror terms, play minimal roles in the overall current drive (as demonstrated in Fig. \ref{fig:QFlat_s01_djdtRHS_testCase}).
It is important to note that electromagnetic effects are not needed for this current drive mechanism.
However, as we will see later, they are required to see the turbulent current's effects on plasma profiles.

The parallel current drive mechanism identified here is universal, appearing in all gyrokinetic simulations that incorporate kinetic electrons.
In conditions where self-interaction is weak, this mechanism leads to fluctuating parallel currents. 
Conversely, when self-interaction is strong --- such as around rational surfaces within a low magnetic shear region --- stationary parallel current corrugations can be maintained. 

\subsection{Turbulent current effects on safety factor profile} \label{sec:turb_curr_eff_q_profile}

In the previous subsection, we have outlined how the radial inhomogeneity of turbulence can drive steady zonal parallel currents.
These in turn will generate a steady zonal parallel component of the magnetic vector potential $A_{\parallel}$ according to Amp\`{e}re's law in Eq. \eqref{eq:Amperes_law}.
When $\beta \neq 0$, this results in a time-stationary zonal perpendicular magnetic field, which in turn changes the safety factor profile.
A detailed derivation of how a steady zonal $A_{\parallel}$ modifies the $q$ profile is presented in \ref{app:turb_mod_q}.

It can be incorporated as an additional term in Eq. \eqref{eq:total_imposed_q} expressed as
\begin{equation}
\label{eq:q_Apar_final}  
    \Tilde{q}_{A_{\parallel}} (x) = \frac{1}{2 \pi} \int_{0}^{2\pi} \frac{\partial \langle A_{\parallel} \rangle_{y,t}}{ \partial x} \frac{J_{xyz} \sqrt{\gamma_{1}}}{C_{xy} C_{y} } dz
\end{equation}
where $\gamma_{1}=g^{xx}g^{yy}-g^{xy}g^{xy}$ is a combination of metric coefficients and $\langle ... \rangle_{t}$ is the time average over the quasi-steady turbulent state.

Taking $\Tilde{q}_{A_{\parallel}}(x)$ into account leads to a new modified form of the safety factor profile
\begin{equation}
\label{eq:total_mod_q}
    q_{tot} = q_{0}+ \frac{q_{0}}{r_{0}} \hat{s}_{0} x + \Tilde{q}(x)+\Tilde{q}_{A_{\parallel}}(x)+ \Delta q,
\end{equation}
which combines the imposed linear and non-uniform safety factor profile terms with a turbulent generated contribution. 
The adjustment of the perpendicular magnetic field, here expressed in terms of $\Tilde{q}_{A_{\parallel}}(x)$, changes the parallel streaming term because particles now follow the altered magnetic field lines instead of the original background field.
This effect enters the gyrokinetic equation through the nonlinear term.

Additionally, we will define the total magnetic shear in a similar way to $q_{tot}$ as 
\begin{equation}
    \hat{s}_{tot} = \hat{s}_{0}+ \Tilde{s}(x)+\Tilde{s}_{A_{\parallel}}(x)
\end{equation}
by differentiating Eq. \eqref{eq:total_mod_q} with respect to the radial coordinate.
Here $\Tilde{s}(x) = (r_{0}/q_{0}) d\Tilde{q}(x)/dx$ is the imposed non-uniform magnetic shear and $\Tilde{s}_{A_{\parallel}}(x)=(r_{0}/q_{0})d\Tilde{q}_{A_{\parallel}}(x)/dx$ is the magnetic shear arising due to stationary zonal $A_{\parallel}$.

\section{Simulation parameters}

For our numerical study, we used both the standard version of the GENE code \cite{Jenko2000, Gorler2011} and the version including the non-uniform safety factor profiles \cite{Ball2022}. 
The principle results presented in this work were obtained through weakly electromagnetic collisionless simulations using the local Miller equilibrium \cite{Miller1998}. 
We primarily considered kinetic electrons as strong turbulence self-interaction results from the non-adiabatic dynamics of the electrons \cite{Dominski2015, Ajay2020, Volcokas2023}. 
However, when instructive, a comparison with adiabatic electrons was made. 

As mentioned earlier, the focus of this work is on the study of standard ion temperature gradient (ITG) turbulence. 
To this end, we employed two different sets of driving gradients when appropriate: the Cyclone Base Case (CBC) \cite{Dimits2000}, which has finite density and temperature gradients for all particle species ($\omega_{T,i} = \omega_{T,e} = 6.96$ and $\omega_{N}=2.22$), and a pure ITG case (referred to as pITG), which has zero density and electron temperature gradients ($\omega_{T,i} =6.96$ and $\omega_{T,e} =\omega_{N}=0$).
Here $\omega_{T,i}=-(R/T_{i})(dT_{i}/dx)$ is the normalized ion temperature gradient, $\omega_{T,e}=-(R/T_{e})(dT_{e}/dx)$ is the normalized electron temperature gradient and $\omega_{N}=-(R/N)(dN/dx)$ is the normalized density gradient, where $R$ is the major radius.
We found that CBC gradients often lead to a mixed ITG and Trapped Electron Mode (TEM) turbulent regime when $\hat{s}_{0} \approx 0$.
Importantly, we keep the plasma $\beta$ small enough, such that electrostatic turbulence always remains dominant over electromagnetic turbulence.

Finally, the simulation parameters for the global particle-in-cell simulations using the ORB5 code are briefly discussed in Sec. \ref{sec:global_sims} alongside the numerical results. 
A more detailed discussion is available in the companion paper \cite{DiGiannatale2024_GlobalReversedShear}.

\section{Numerical results}

\subsection{Turbulent corrugations of linear safety factor profiles}\label{sec:corrug_linear_profile}
We start by studying how linear safety factor profiles are modified by the currents generated by weakly electromagnetic turbulence.
We will explore scenarios with low, but finite average magnetic shear, (specifically $|\hat{s}_{0}|=0.1$) using CBC and pITG kinetic electron simulations.
We demonstrate that when magnetic shear is small, the zonal component of $A_{\parallel}$ can significantly alter the safety factor profile, resulting in a stepped profile with flattening at different low order rational surfaces that in turn leads to a substantial reduction of turbulent transport. 
It is crucial to note that the main results were achieved using the standard version of the GENE code. 
Simulations with non-linear safety factor profiles will be addressed in later sections.

\begin{figure}[!htb]
\centering
\includegraphics[width=0.85\textwidth]{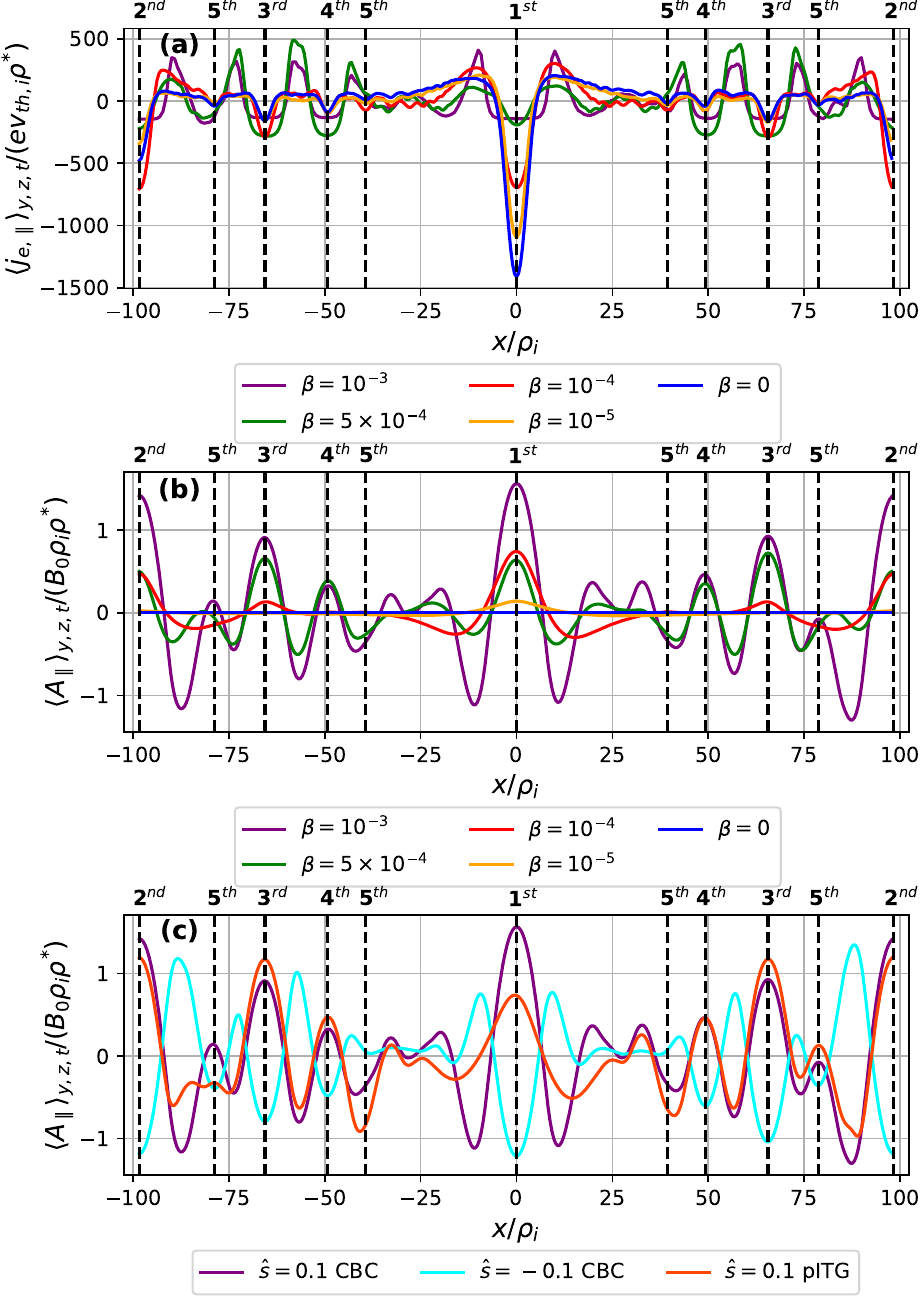}
\caption{The radial profile of flux surface and time averaged (a) parallel electron current and (b) magnetic potential for different values of $\beta \in \{0, 10^{-5}, 10^{-4}, 5\times 10^{-4}, 10^{-3}\}$, fixed $\hat{s}_{0}=0.1$, and fixed CBC gradients. (c) Magnetic potential for different magnetic shear $\hat{s}_{0}$ and gradients values (CBC and pITG) at fixed $\beta = 10^{-3}$. The black vertical dotted lines indicate low order rational surfaces up to $5^{th}$ order. Except for the pITG case, the simulations were performed using the parameters shown in Table \ref{tab:parameters_nonlinear_linQ_s01_highres} of \ref{app:simulation_parameters}.}
\label{fig:QFlat_s01_Apar_betaScan_070524}
\end{figure}

Corrugations in radial plasma profiles, such as temperature, density, and shearing rate, around low order rational surfaces have been observed and studied in detail in both local \cite{Dominski2015, Ball2019, Ajay2020, Volcokas2024_PartTwo_NL} and global \cite{Waltz2006, Dominski2017} gyrokinetic simulations. 
Similarly, we find stationary zonal average parallel velocity (equivalently parallel charge current) profile corrugations due to turbulent current drive around rational surfaces, as shown in Fig. \ref{fig:QFlat_s01_Apar_betaScan_070524}(a).
This leads to stationary magnetic potential $A_{\parallel}$ corrugations according to Amp\`{e}re's law in Eq. \eqref{eq:Amperes_law}, which are shown in Fig. \ref{fig:QFlat_s01_Apar_betaScan_070524}(b) for CBC $\hat{s}_{0}=0.1$ simulations with varying $\beta$ values\footnote{It is a common practice in gyrokinetic simulations to use small $\beta \leq 10^{-4}$ to avoid the electrostatic shear Alfven wave and speed up numerical simulations.
We emphasize that even these low values of $\beta$ can enable the turbulent-current feedback mechanisms studied in this work, particularly if magnetic shear is low. 
This is especially true for stellarators where it is much more common for the global magnetic shear tends to be low.}.
The amplitude of $\langle A_{\parallel} \rangle_{y,z,t}$  depends on $\beta$ and becomes more prominent around low order rational surfaces as $\beta$ is increased.
As discussed in Sec. \ref{sec:turb_curr_eff_q_profile} and as we will see later on, the zonal component of $A_{\parallel}$ leads to safety factor profile flattening at rational surfaces.
We also performed simulations with negative average magnetic shear $\hat{s}_{0}=-0.1$.
These exhibit similar magnetic potential corrugations, but the corrugations have the opposite sign to positive shear cases (as shown in Fig. \ref{fig:QFlat_s01_Apar_betaScan_070524}(c)), which implies that the safety factor profile is flattened at rational surfaces regardless of the sign of $\hat{s}_{0}$.
Also in Fig. \ref{fig:QFlat_s01_Apar_betaScan_070524}(c), we see that in a pITG simulation, similar magnetic potential corrugations also build up, regardless of the very different gradients driving the turbulence.
For the remainder of this section, we will focus on the less physically idealized CBC simulations. 
Finally, magnetic potential corrugations have also been observed in simulations with more standard values of the magnetic shear $\hat{s}_{0} \sim 1$, but, given the larger $\hat{s}_{0}$, they are unable to significantly impact the imposed linear safety factor profile.

\begin{figure}[!hb]
\centering
\includegraphics[width=0.85\textwidth]{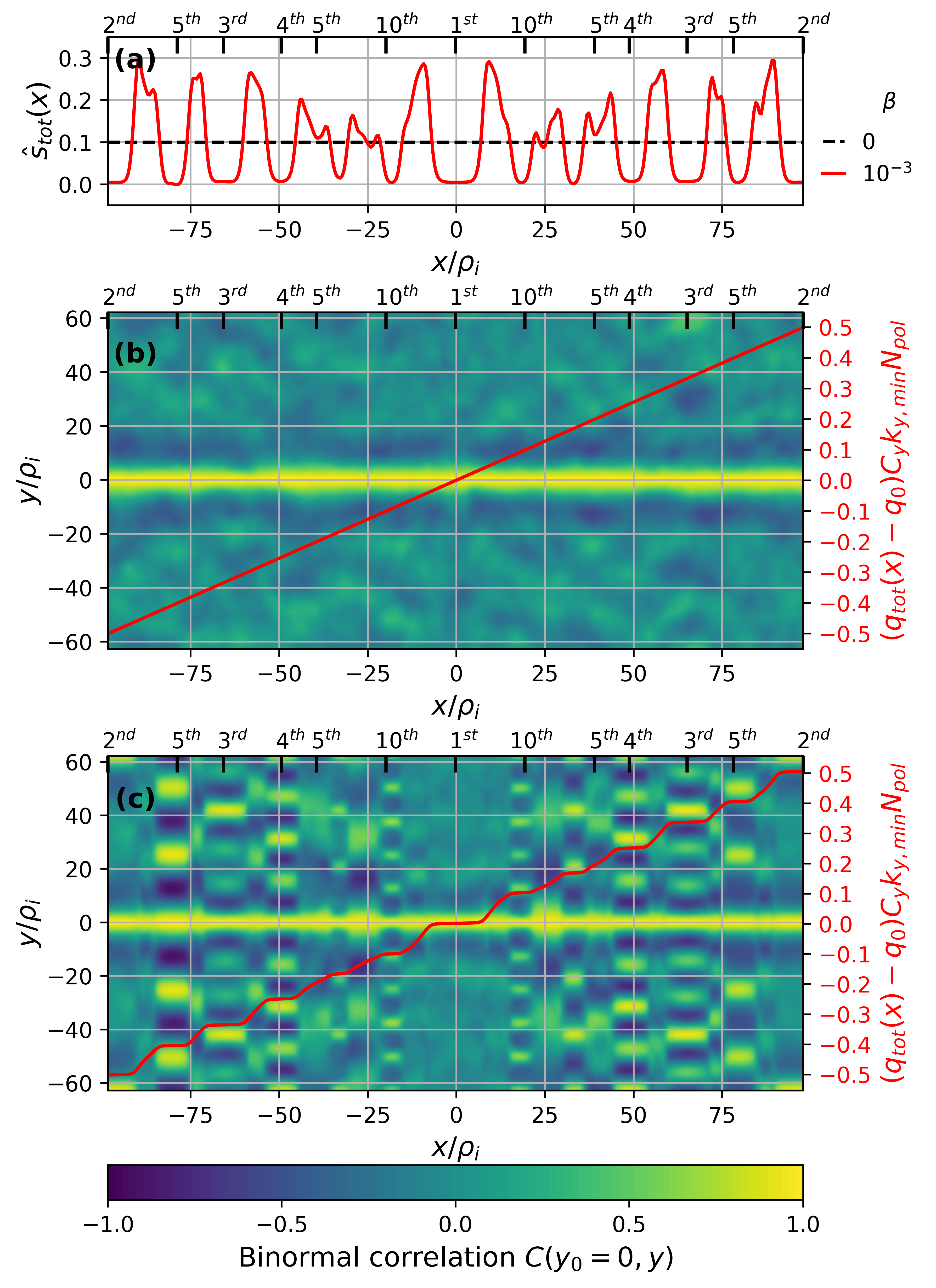}
\caption{(a) Radial flux-surface-averaged magnetic shear profile for electrostatic ($\beta=0$) and weakly electromagnetic ($\beta=10^{-3}$) simulations, with a fixed background magnetic shear $\hat{s}_{0}=0.1$. The lower subplots display the binormal electrostatic potential auto-correlation $C(y_{0}=0, y)$ for $z=0$ (color plot) as a function of radial position $x$ as well as the safety factor profile plotted as $(q_{tot}(x)-q_{0})C_{y}k_{y,min}N_{pol}$ (red line) for (b) purely electrostatic ($\beta=0$) and (c) weakly electromagnetic ($\beta = 10^{-3}$) simulations. The top axes show the order and positions of selected rational surfaces. The simulations were performed using parameters shown in Table \ref{tab:parameters_nonlinear_linQ_s01_highres} of \ref{app:simulation_parameters}.}
\label{fig:QFlat_s01_biCorr_b0vsb0001_paper_160524}
\end{figure}

The effective magnetic shear profiles from the CBC electromagnetic $\beta=10^{-3}$ and electrostatic $\beta = 0$ simulations are compared in Fig. \ref{fig:QFlat_s01_biCorr_b0vsb0001_paper_160524}(a). 
The safety factor profiles for the two cases are shown in Fig. \ref{fig:QFlat_s01_biCorr_b0vsb0001_paper_160524}(b) and (c) respectively, where a stepped safety factor profile can be easily seen for the electromagnetic case.
Note that the right-hand side vertical axis represents the value of the safety factor profile, normalized in such a way that the value represents the inverse of the order of the rational magnetic surfaces.
For example, the value of $(q_{tot}(x)-q_{0})C_{y}k_{y, min}N_{pol} = 0.5$ would indicate a second order rational surface, while $(q_{tot}(x)-q_{0})C_{y}k_{y, min}N_{pol} = 0$ corresponds to the lowest order rational surface (see related discussion of parameter $\eta$ in Eq. \eqref{eq:phase_factor_C}).

At radial locations where the safety factor profile is flattened, the characteristics of turbulent eddies undergo significant changes.
Indeed, when effective magnetic shear is zero, turbulent eddies can extend along the magnetic field lines for hundreds of poloidal turns around the torus, enhancing turbulent self-interaction.
One of the consequences of this stronger self-interaction is ``eddy squeezing" at certain radial locations, which is known to lead to reduced turbulent transport \cite{Volcokas2023, Volcokas2024_PartTwo_NL}.
This effect can be visualized by comparing binormal correlation $C(y_{0}=0, z_{0}=0, y, x)$ at the outboard midplane $z_{0}=0$ as a function of the radial coordinate between the two simulations in Fig. \ref{fig:QFlat_s01_biCorr_b0vsb0001_paper_160524}(b) and (c).
The correlation function is defined in terms of two-point Eulerian correlation \cite{Wallace2014}
\begin{equation}
\label{eq:correlation_def}
    C(\mathbf{x_{1}}, \mathbf{x_{2}}) = \frac{\langle \phi_{NZ}(\mathbf{x_{1}},t)\phi_{NZ}(\mathbf{x_{2}},t)\rangle_{t}}{\sqrt{\langle \phi^{2}_{NZ}(\mathbf{x_{1}},t)\rangle_{t}}\sqrt{\langle \phi^{2}_{NZ}(\mathbf{x_{2}},t)\rangle_{t}}},
\end{equation}
where $\mathbf{x_{1}}$ and $\mathbf{x_{2}}$ are two real space locations between which the electrostatic potential correlation is measured, $\phi_{NZ}=\phi-\langle \phi \rangle_{y}$ is the non-zonal component of electrostatic potential, and $\langle ... \rangle_{t}$ is the temporal average over the turbulence scale.
For the electromagnetic case, we can easily identify $1^{st}$, $2^{nd}$, $3^{rd}$, $4^{th}$, $5^{th}$ and even $10^{th}$ order rational surfaces by the number of peaks in the binormal direction.
This corresponds to the number of times turbulent eddies pass through the parallel boundary before ``biting its own tail".
For example, when an eddy stretches through the parallel boundary on a 5$^{th}$ order surface, it is shifted by a fifth of the box width.
In the case for $\beta = 0$, the binormal correlation profile is essentially the same for all positions along the radial domain with only minor variations at the lowest order rational surfaces, i.e. showing no strong indication of long eddies biting their own tail at low order rational surfaces.
Therefore, the corrugated pattern in the correlation and the stepped safety factor profile for the weakly electromagnetic case show that turbulence-generated currents have a profound effect on the imposed safety factor profile when magnetic shear is low.

\begin{figure}[!hb]
\centering
\includegraphics[width=0.85\textwidth]{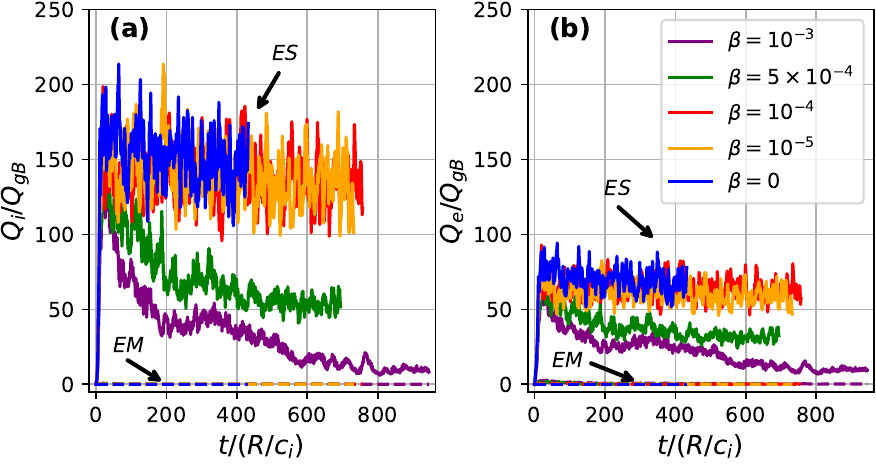}
\caption{Time traces of (a) ion and (b) electron electrostatic and electromagnetic heat fluxes for CBC drive at fixed $\hat{s}_{0}=0.1$ with varying $\beta$. The simulations were performed using the parameters shown in Table \ref{tab:parameters_nonlinear_linQ_s01_highres} of \ref{app:simulation_parameters}.}
\label{fig:QFlat_s01_Qtrace_betaScan_080524}
\end{figure}

\begin{figure}[!hb]
\centering
\includegraphics[width=0.85\textwidth]{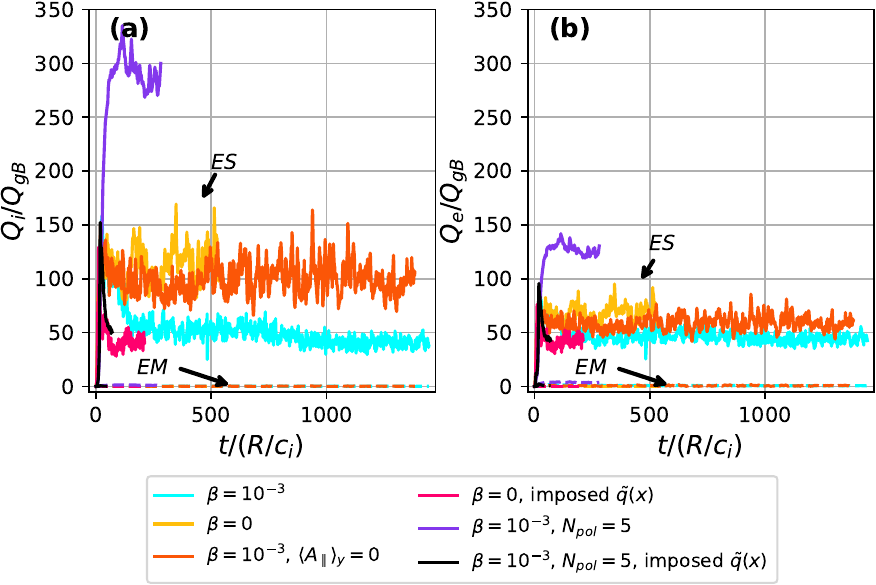}
\caption{Time traces of (a) ion and (b) electron electrostatic and electromagnetic heat fluxes for CBC drive at fixed $\hat{s}_{0}=0.1$ and heavier electrons $ m_{i}/m_{e} = 184$ with either $\beta = 0$ or $\beta = 10^{-3}$ and zonal $\langle A_{\parallel}\rangle_{y}$ either kept or set to zero. The simulations were performed using the parameters shown in Table \ref{tab:parameters_nonlinear_linQ_s01_lowres_mex20} of \ref{app:simulation_parameters}.}
\label{fig:QFlat_s01_mex50_Qtrace_betaScan_080524}
\end{figure}

\begin{figure}[!hb]
\centering
\includegraphics[width=0.55\textwidth]{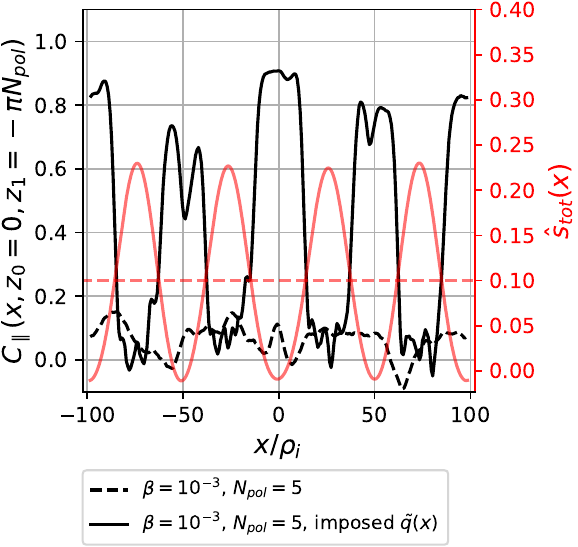}
\caption{The parallel correlation of the electrostatic potential between the outboard midplane $z_{0}=0$ and parallel boundary $z_{1}=- \pi N_{pol}$ measured along the magnetic field lines as a function of radial position $x$ (black) as well as the corresponding magnetic shear profile (red) for the same $N_{pol}=5$ simulations as in Fig \ref{fig:QFlat_s01_mex50_Qtrace_betaScan_080524}. Results for the simulation without (respectively with) imposed $\Tilde{q}(x)$ modulation are represented by dashed (respectively full) lines. The simulations were performed using the parameters shown in Table \ref{tab:parameters_nonlinear_linQ_s01_lowres_mex20} of \ref{app:simulation_parameters}.}
\label{fig:QFlat_s01_mex50_Npol5_CorrComp}
\end{figure}

Crucially, the modification of the safety factor profile by the zonal component of $A_{\parallel}$ results in a stabilization of turbulence, as depicted in Fig. \ref{fig:QFlat_s01_Qtrace_betaScan_080524}, where significantly reduced ion and electron heat fluxes are observed at $\beta \sim 10^{-3}$.
We can attribute this reduction in transport primarily to zonal $A_{\parallel}$, which we will justify using Figs. \ref{fig:QFlat_s01_mex50_Qtrace_betaScan_080524} and \ref{fig:QFlat_s01_mex50_Npol5_CorrComp}. Fig. \ref{fig:QFlat_s01_mex50_Qtrace_betaScan_080524} compares an electrostatic (ES) simulation with an electromagnetic (EM) case where the zonal $A_{\parallel}$ is artificially set to zero (compare yellow and orange curves).
The EM simulation without zonal $A_{\parallel}$ yields heat fluxes similar to those in the ES case.
The opposite is also true --- when a non-uniform safety factor profile, mimicking the effects of zonal $A_{\parallel}$ in EM scenarios, is imposed onto ES simulations, transport is again significantly reduced (compare cyan and magenta curves in Fig. \ref{fig:QFlat_s01_mex50_Qtrace_betaScan_080524}).
To make this comparison the stepped safety factor profile $\Tilde{q}(x)$ was taken from the standard $\beta = 10^{-3}$ case in Fig. \ref{fig:QFlat_s01_mex50_Qtrace_betaScan_080524} and imposed on an ES simulation using the non-uniform shear functionality expressed by Eq. \eqref{eq:non_uniform_q}.

Additionally, a simulation with $N_{pol}=5$ was performed to eliminate parallel self-interaction.
Without self-interaction a significant increase in heat flux is observed compared with the $N_{pol}=1$ simulations, which further confirms the stabilizing effect of self-interaction and resulting modifications to the safety factor profile. 
However, if an $N_{pol}=5$ simulation is started with an imposed non-uniform safety factor profile to mimic the effects of zonal $A_{\parallel}$ in the $N_{pol}=1$ case, the heat flux is reduced by about four times.
This reduction is due to the imposed zero magnetic shear regions, which allow much longer turbulent eddies to form and facilitate strong parallel self-interaction, even when $N_{pol} = 5$.
The increased turbulent eddy length in the $\hat{s}_{0} \approx 0$ regions is illustrated in Fig. \ref{fig:QFlat_s01_mex50_Npol5_CorrComp}, which shows the parallel correlation $C_{\parallel}(x)=\langle C(z_{0}, z_{b}, x, y) \rangle_{y}$ as defined in Eq. \eqref{eq:correlation_def} between the central outboard midplane $z_{0}=0$ and the parallel boundary $z_{b}=\pi N_{pol}$.
While the $N_{pol} = 5$ simulation with an imposed non-uniform safety factor profile was performed for a short time due to its high computational cost, we believe the trends are clear and agree with the previous results.
These observations underscore the critical role of zonal $A_{\parallel}$ in reducing turbulent transport by modifying the safety factor profile to enhance the effect of rational surfaces and low $\hat{s}_{0}$.

It should be noted that the simulations shown in Figs. \ref{fig:QFlat_s01_mex50_Qtrace_betaScan_080524} and \ref{fig:QFlat_s01_mex50_Npol5_CorrComp} were conducted at lower phase space resolution and with twenty times higher electron mass compared to those in Fig. \ref{fig:QFlat_s01_Qtrace_betaScan_080524}. 
This adjustment was necessary to reduce the increased computational cost due to the longer computational domain ($N_{pol}=5$) and non-uniform safety factor profile modulation ($\Tilde{q}(x) \neq 0$).

\begin{figure}[!hb]
\centering
\includegraphics[width=0.85\textwidth]{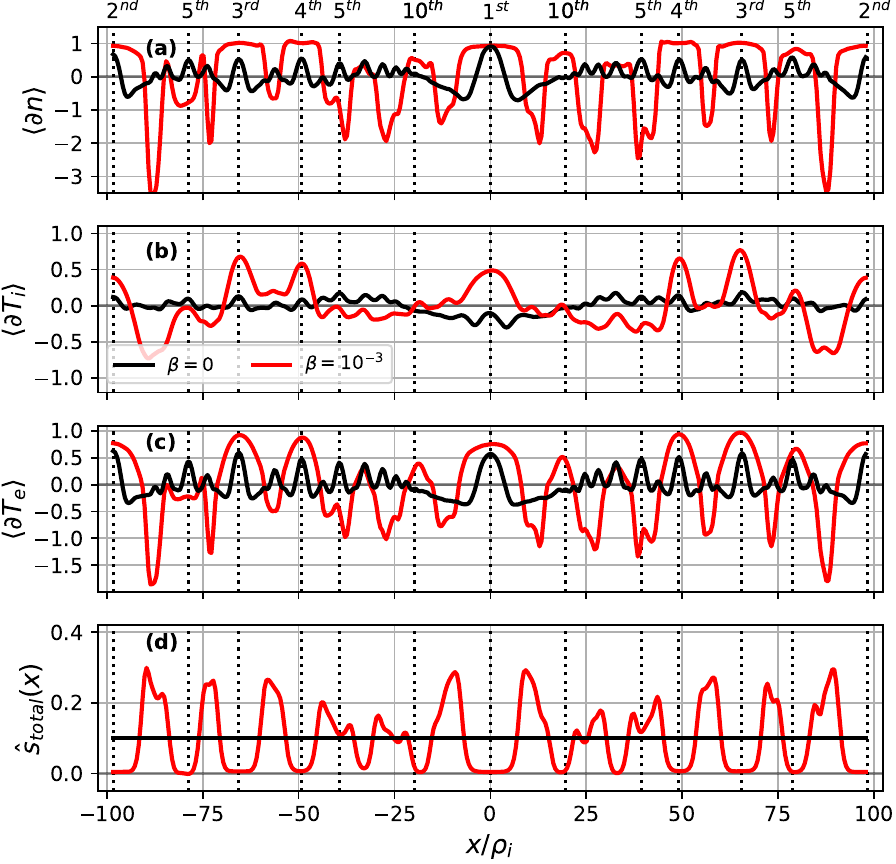}
\caption{The radial profile of $\langle \partial A \rangle = - \frac{d \langle \delta A \rangle}{dx}/ \frac{d A}{d x} $ where $A$ is the (a) density, (b) ion temperature or (c) electron temperature for $\hat{s}=0.1$ simulations with varying $\beta$. The final subplot (d) shows an effective radial magnetic shear profile. The black vertical dotted lines indicate different order integer surfaces. The simulations were performed using the parameters shown in Table \ref{tab:parameters_nonlinear_linQ_s01_highres} of \ref{app:simulation_parameters}.}
\label{fig:QFlat_s01_b0vsb0001_ProfileCorrugations_300524}
\end{figure}

In addition to a strong stabilization effect due to safety factor profile corrugations from $A_{\parallel}$, we also observe significant changes to other plasma profiles.
Specifically, the stationary corrugations in density and temperature become large relative to their background gradients. 
The total radial profile $\Bar{A}(x)$ of a physical quantity $A$ (e.g. density $n$, temperature $T$, flow $u_{\parallel}$) modulated by the turbulence is
\begin{equation}
    \Bar{A}=A_{0}+\frac{d A}{d x}x +\langle \delta A \rangle,
\end{equation}
where $A_{0}$ is the imposed background value, $d A / d x$ is the imposed background gradient, $\langle \delta A \rangle$ stands for the flux surface and time average of the fluctuating quantity $\delta A$
\begin{equation}
    \langle \delta A \rangle = \langle \delta A(x,y,z,t) \rangle_{y,z,t}.
\end{equation}
The effective gradient is then
\begin{equation}
    \frac{d\Bar{A}}{dx} = \frac{d A}{d x} + \frac{d \langle \delta A \rangle}{d x}.
\end{equation}
This is similar to the total safety factor and magnetic shear profiles in Eq. \eqref{eq:qtotal_final_app}, however, without additionally imposed non-uniformity. 
Thus, the ratio of the corrugation gradient to the background gradient is defined as \cite{Volcokas2024_PartTwo_NL}
\begin{equation}
    \left<\partial A \right> = - \frac{\frac{d}{dx}\langle \delta A \rangle}{\frac{d A}{d x}}.
\end{equation}
If $\left<\partial A \right>$ is positive, the plasma profile is locally flattened (assuming $dA/dx<0$), and if it is negative, the profile is steepened. The value $\langle \partial A \rangle = 1$ represents the complete flattening of the corresponding plasma quantity.
This ratio is shown alongside the radial magnetic shear profiles for the electrostatic and electromagnetic simulations in Fig. \ref{fig:QFlat_s01_b0vsb0001_ProfileCorrugations_300524}. 

While the main observed effect of the stepped safety factor profile is a significant reduction in turbulent transport and turbulence stabilization, once a quasi-steady state is reached, the radial regions with zero magnetic shear facilitate more efficient transport compared to higher magnetic shear regions.
This leads to flattened effective density and temperature gradients in the zero magnetic shear regions around $1^{st}$, $2^{nd}$, $3^{rd}$ and $4^{th}$ order rational surfaces, as seen in Fig. \ref{fig:QFlat_s01_b0vsb0001_ProfileCorrugations_300524} for the electromagnetic case.
Since the average gradient across the whole box is fixed (flux tube simulations being effectively gradient driven), this leads to steepening in the profiles in the high shear radial regions.

The study suggests that stabilization primarily arises from the turbulence-modified $q$ profile and we wish to understand how. 
Contrary to what might be initially assumed, we will see that it results from increased self-interaction from longer eddies emerging in the $\hat{s} \approx 0$ regions, rather than from direct effects of profile curvature or higher magnetic shear regions.

\begin{figure}[!hb]
\centering
\includegraphics[width=0.85\textwidth]{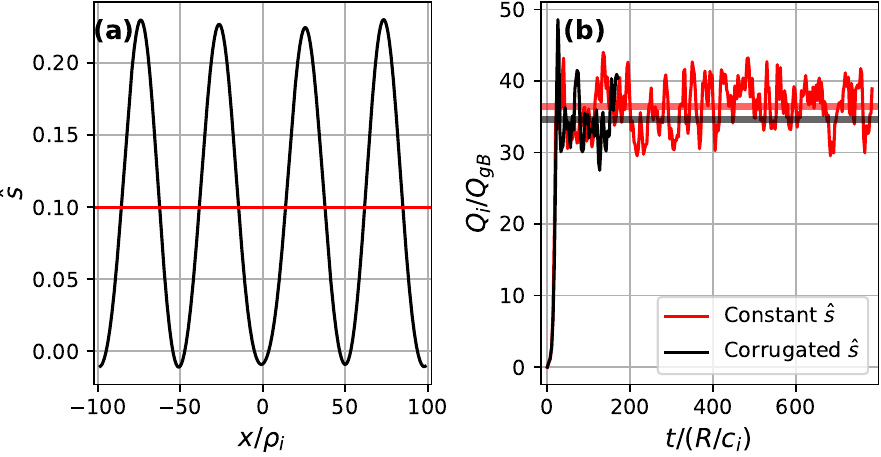}
\caption{The (a) total magnetic shear profile and (b) corresponding time traces of the electrostatic ion heat flux for CBC drive with background magnetic shear $\hat{s}_{0}=0.1$ and adiabatic electrons. The thick horizontal lines in (b) indicate the heat flux values time-averaged over the quasi-steady turbulent phase. The simulations were performed using the parameters shown in Table \ref{tab:parameters_nonlinear_linQ_s01_lowres_mex20} of \ref{app:simulation_parameters}.}
\label{fig:QFlat_s01_AE_linearVSsteppedQ_051424}
\end{figure}

To eliminate strong parallel eddy self-interaction and isolate the effects of safety factor profile curvature on the heat flux, adiabatic electron simulations were carried out.
Kinetic electrons could be used if the domain is sufficiently long, but this proved too computationally expensive.
A comparison of the magnetic shear profiles and heat flux time traces from adiabatic simulations is presented in Fig. \ref{fig:QFlat_s01_AE_linearVSsteppedQ_051424}. 
One simulation has a standard linear $q$ profile, while the other has an imposed safety factor profile with a pronounced corrugation, taking the profile from kinetic electron simulations.
Unlike the simulations shown in Fig. 
\ref{fig:QFlat_s01_mex50_Npol5_CorrComp}, due to the adiabatic treatment of electrons the parallel eddy length does not increase in zero-shear regions, keeping parallel self-interaction negligible.
We find that the imposed magnetic shear non-uniformity has no substantial effect on the turbulent heat flux, nor the radial size of the turbulent eddies (based on characteristic eddy sizes obtained from the auto-correlation of electrostatic potential).
The perpendicular eddy size is similar between kinetic and adiabatic electron simulations, suggesting that the direct impact of the corrugations on radial turbulence in kinetic simulations via safety factor profile curvature would be similarly unimportant.

\begin{figure}[!hb]
\centering
\includegraphics[width=0.55\textwidth]{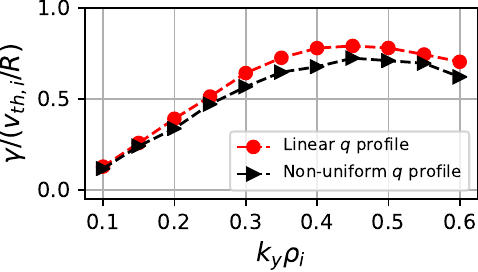}
\caption{The linear growth rate dependence as a function of the binormal wave number $k_{y}$ for kinetic electron simulations considering either a linear $q$ profile (red dots) or a non-uniform $q$ profile (black triangles) with the non-uniform $q$ profile shown in Fig. \ref{fig:QFlat_s01_AE_linearVSsteppedQ_051424} (a). The key simulation parameters are $\hat{s}_{0}=0.1$, $\Delta y_{0} = 0$, $k_{x} \rho_{i} = 0$, $L_{x} \approx 200 \rho_{i}$ and CBC gradients. }
\label{fig:Qflat_s01_NonUniform_gamma}
\end{figure}

To add further evidence, we also performed linear simulations with kinetic electrons including or not the safety factor profile modulations.
No significant change in the linear growth rate was observed, in agreement with the nonlinear adiabatic results.
The comparison of the growth rate as a function of the binormal wavenumber between linear and non-uniform safety factor profiles is shown in Fig. \ref{fig:Qflat_s01_NonUniform_gamma}.

Consequently, we attribute the significant reduction of turbulent transport observed in the nonlinear kinetic electron simulations shown in Figs. \ref{fig:QFlat_s01_Qtrace_betaScan_080524} and \ref{fig:QFlat_s01_mex50_Qtrace_betaScan_080524} primarily to strong parallel turbulent self-interaction at rational surfaces within broad zero magnetic shear regions. 
Neither perpendicular turbulence dynamics nor linear physics seem to be significantly affected by the modulation of the safety factor profile and thus do not explain the observed reduction in the heat flux.

Finally, while magnetic islands are present in the simulations shown in Figs. \ref{fig:QFlat_s01_Qtrace_betaScan_080524} and \ref{fig:QFlat_s01_mex50_Qtrace_betaScan_080524}, we believe they play a secondary role to the zonal $A_{\parallel}$ component.
Magnetic islands are related to a stationary non-zonal component of $A_{\parallel}$, are localized around rational surfaces, and have even parity.
First, the amplitudes of the non-zonal $A_{\parallel}$ components are an order of magnitude smaller than their zonal counterpart. 
Second, as demonstrated in Fig. \ref{fig:QFlat_s01_mex50_Qtrace_betaScan_080524}, either artificially suppressing zonal $A_{\parallel}$ or imposing safety factor corrugations results in the most significant impact on heat flux.
Comparing the cyan and magenta curves in Fig. \ref{fig:QFlat_s01_mex50_Qtrace_betaScan_080524} isolates the impact of non-zonal $A_{\parallel}$, showing that it is small.

In summary, the findings from this section demonstrate an enhanced self-regulation mechanism of turbulence through a back reaction on the safety factor profile. 
It leads to significant corrugations of the plasma profiles and, most strikingly, to a stepped safety factor profile, which strongly amplifies the stabilizing effects of self-interaction on turbulent transport.

\subsection{Turbulent flattening of small perturbations to a flat safety factor profile}\label{sec:flat_small_perturb}
We have shown that when $\hat{s}_{0}$ is small, the imposed safety factor profile in standard flux tube simulations can be modified by turbulent-generated currents, leading to fully flattened regions around low-order rational surfaces.
To explore this effect further, we will study how turbulence responds to weak (sinusoidal) modulations of a safety factor profile with zero average magnetic shear.
This is the simplest possible scenario and will allow us to better understand the system's reaction to safety factor profile perturbations.

\begin{figure}[!hbt]
\centering
\includegraphics[width=0.65\textwidth]{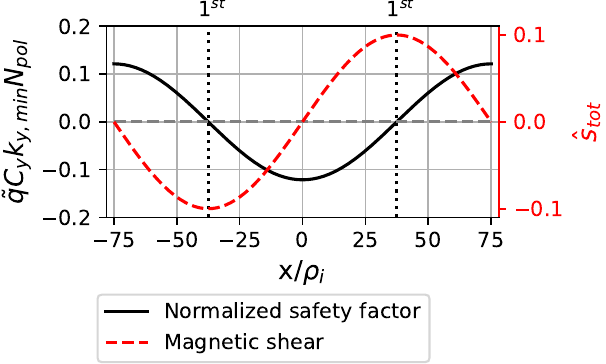}
\caption{The imposed non-uniform safety factor profile (black solid line) and the corresponding magnetic shear profile (red dashed line), plotted as functions of the radial position. These profiles were generated using a single Fourier sine coefficient,  $\hat{s}^{S}_{1} = 0.1$, with an average magnetic shear of $\hat{s}_{0} = 0$. Black vertical dotted lines mark the locations of first-order rational surfaces.}
\label{fig:QFlat_sSm01_example}
\end{figure}

Unless noted otherwise, the simulations discussed in this section use a background magnetic shear of $\hat{s}_{0}=0$, and no background binormal shift $\Delta y_{0} = 0$. 
We introduce a weak magnetic shear non-uniformity by including a single lowest harmonic Fourier sine coefficient $\hat{s}^{S}_{1}$ in Eq. \eqref{eq:non_uniform_q}.
Thus we are specifying a magnetic shear profile that is a sine wave --- or equivalently, a safety factor profile that is a cosine. 
By ``weak non-uniformity" we mean that the Fourier coefficient $\hat{s}^{S}_{1}$ is sufficiently small, such that no low-order rational surface (besides the lowest-order rational surfaces) is present in our simulation domain and can facilitate significant turbulent self-interaction.
To further ensure that self-interaction at other low-order rational surfaces is avoided, these simulations were performed with heavy electrons $m_{i}/m_{e} = 368$, which reduces parallel eddy extent.
In the simulation domain, only two low-order rational surfaces, in this case, $1^{st}$ order integer surfaces, are present, as shown in Fig. \ref{fig:QFlat_sSm01_example}.

\begin{figure}[!hbt]
\centering
\includegraphics[width=0.85\textwidth]{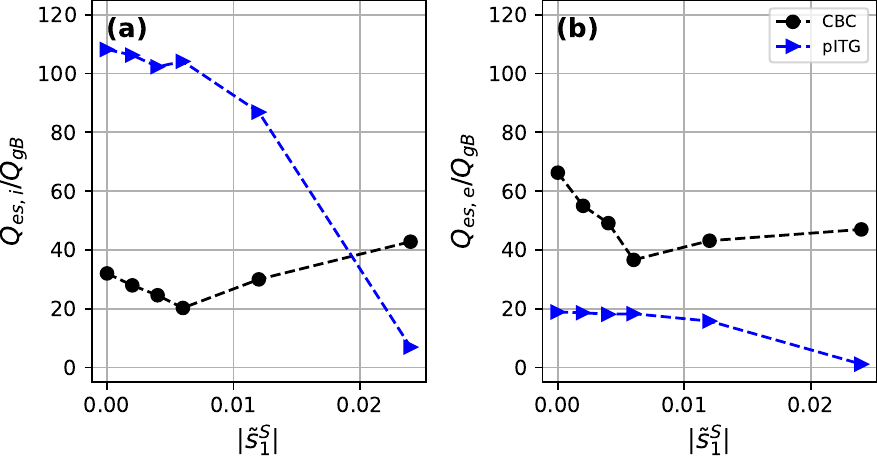}
\caption{Dependence of the (a) ion and (b) electron heat fluxes on the amplitude $|\tilde{s}^{S}_{1}|$ of the magnetic shear for CBC-like (black circles) and pITG (blue triangles) parameters. The average magnetic shear is $\hat{s}_{0}=0$, $\beta=0$, and other parameters are given in Table \ref{tab:parameters_nonlinear_sSsmall_study} of \ref{app:simulation_parameters}.}
\label{fig:Qflat_Q_vs_sS_pITGandCBC_300424}
\end{figure}

The electrostatic simulations with this imposed magnetic shear non-uniformity display a behavior similar to those discussed in our previous work in Refs. \cite{Volcokas2023, Volcokas2024_PartOne_L, Volcokas2024_PartTwo_NL}.
In these past studies, we varied the parallel background boundary shift $\Delta y_{0}$ in simulations with \textit{constant} background magnetic shear $\hat{s}_{0} = 0$ and without magnetic shear non-uniformity.
It was observed that different values of the boundary shift or, in other words, different magnetic topologies significantly impact turbulence stability, sometimes leading to complete turbulence stabilization.
Unlike in these previous works, in the simulations presented here, the magnetic field line topology varies radially due to the safety factor non-uniformity leading to a small radially dependent binormal offset $\Delta y (x) = \Tilde{q}(x) C_{y} k_{y,min} N_{pol}$, with $\Tilde{q}(x)$ given by Eq. \eqref{eq:non_uniform_q}.
At all locations apart from the integer surfaces, the field lines experience a small binormal offset when the field lines pass through the parallel boundary condition.
While the shift is not uniform, we expect that the corresponding small field line offset $\Delta y$ on each magnetic surface across the radial domain results in an average topological change effect.
Given the sensitivity of the turbulence to small changes around integer surfaces (see Fig. 7 of \cite{Volcokas2024_PartTwo_NL}), this may be important.
Fig. \ref{fig:Qflat_Q_vs_sS_pITGandCBC_300424} shows the average heat flux as a function of non-uniformity amplitude from purely electrostatic CBC-like and pITG simulations. 
We see that even small modulations of the safety factor profile lead to very significant turbulence stabilization in pITG simulations.
In CBC-like simulations, there is a marginal reduction in transport when $0 < \Tilde{s}^{S}_{1} \leq 5 \times 10^{3}$ and a destabilization when $\Tilde{s}^{S}_{1} \geq 5 \times 10^{3}$, as one would expect from Fig. 7 in \cite{Volcokas2024_PartTwo_NL}. 

\begin{figure}[!hbt]
\centering
\includegraphics[width=0.85\textwidth]{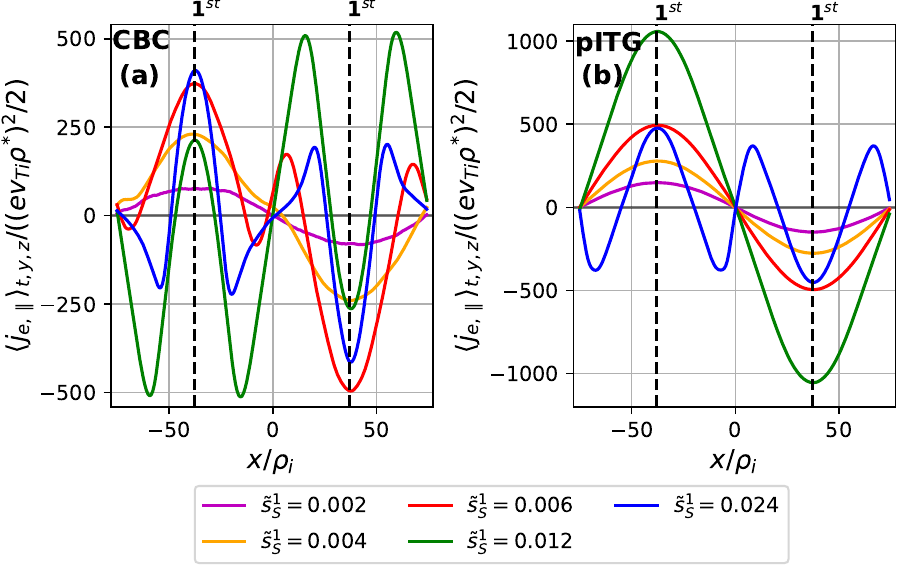}
\caption{Radial profiles of flux surface and time averaged parallel electron current for (a) CBC-like and (b) pITG simulations with varying non-uniform magnetic shear amplitude $\tilde{s}_{1}^{S}$. Black vertical dotted lines mark the locations of first-order rational surfaces. The current is normalized using the thermal ion velocity $v_{T_{i}}$ and gyroradius-to-machine-size ratio $\rho^{*}$. The average magnetic shear is $\hat{s}_{0}=0$, $\beta=0$, and other parameters are given in Table \ref{tab:parameters_nonlinear_sSsmall_study} of \ref{app:simulation_parameters}.}
\label{fig:QFlat_small_sSm_pITGandCBC_upar_sSmScan_300424}
\end{figure}

It is important to note that stationary parallel current layers build up around integer surfaces in these electrostatic simulations as shown in Fig. \ref{fig:QFlat_small_sSm_pITGandCBC_upar_sSmScan_300424}.
This is consistent with the turbulent current generation mechanism discussed in Sec. \ref{sec:turbulent_current_drive} and with the linear safety factor profile simulations in Sec \ref{sec:corrug_linear_profile} (see Fig. \ref{fig:QFlat_s01_Apar_betaScan_070524}).
In both CBC-like and pITG simulations, corrugations in the current profile, generally increase as the non-uniformity amplitude increases.
However, in electrostatic simulations, these turbulence-generated currents are unable to act back on the safety factor profile.

\begin{figure}[!hbt]
\centering
\includegraphics[width=0.5\textwidth]{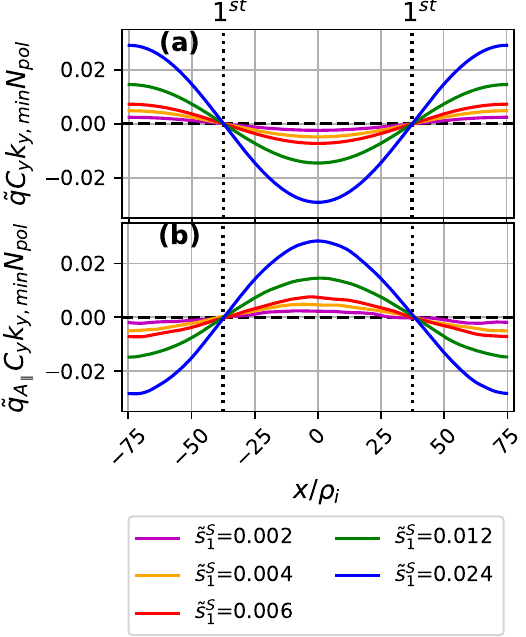}
\caption{(a) Imposed safety factor non-uniformity and (b) safety factor modulation due to turbulent generated zonal $A_{\parallel}$ for a set of CBC-like simulations with varying amplitude of the imposed sinusoidal non-uniformity. Black dotted vertical lines indicate the initial locations of 1$^{st}$ order rational surfaces. The background magnetic shear is $\hat{s}_{0}=0$ and $\beta = 10^{-3}$. Other parameters are given in Table \ref{tab:parameters_nonlinear_sSsmall_study} of \ref{app:simulation_parameters}.}
\label{fig:QFlat_small_sS_perturbation}
\end{figure}

When finite but small $\beta = 10^{-3}$ is introduced, we observe significant changes in the system's behavior. 
We find that turbulent generated stationary zonal currents almost completely cancel out the imposed small amplitude non-uniformity. 
We observe this phenomenon in both CBC and pITG simulations.
For the remainder of this section, we will again focus exclusively on CBC simulations, since they represent a less idealized physical case.
The cancellation, corresponding to a flattening of the safety factor profile, is illustrated in Fig. \ref{fig:QFlat_small_sS_perturbation}, which compares the imposed safety factor non-uniformity $\Tilde{q}(x)$ to the safety factor profile modulations $\Tilde{q}_{A_{\parallel}}(x)$ generated by turbulent driven currents according to Eq. \eqref{eq:q_Apar_final}.
We see that the imposed non-uniformity $\Tilde{q}(s)$ is indeed exactly canceled out by the turbulent-driven modulations $\Tilde{q}_{A_{\parallel}}(x)$.

\begin{figure}[!hbt]
\centering
\includegraphics[width=0.85\textwidth]{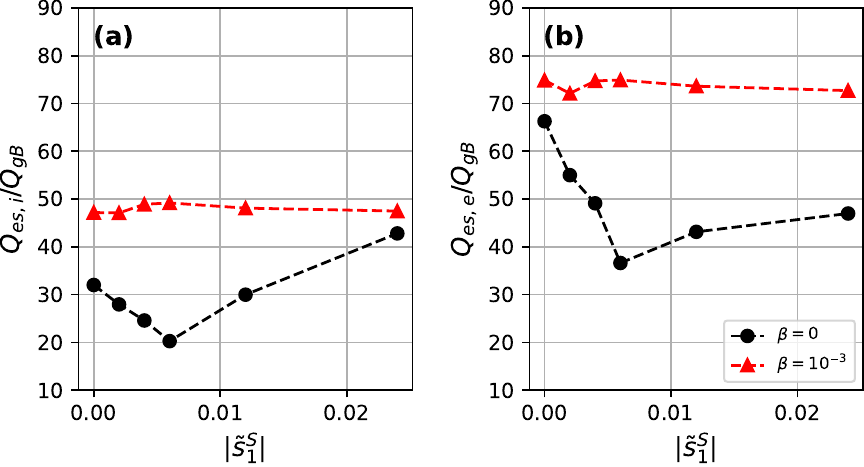}
\caption{The dependence of (a) ion and (b) electron heat fluxes on the amplitude $\Tilde{s}^{S}_{1}$ of non-uniform magnetic shear for a set of CBC-like simulations. Black circles indicate simulations with $\beta=0$ (same as in Fig. \ref{fig:Qflat_Q_vs_sS_pITGandCBC_300424}) and red triangles indicate simulations with $\beta=10^{-3}$. The background magnetic shear is $\hat{s}_{0}=0$ and the other parameters are given in Table \ref{tab:parameters_nonlinear_sSsmall_study} of \ref{app:simulation_parameters}.}
\label{fig:Qflat_Q_vs_sS_beta0vs0001}
\end{figure}

This flattening of the safety factor profile significantly impacts the heat flux. 
As one would expect, the simulations with $\beta \neq 0$ have heat fluxes very close to standard $\hat{s}_{0}=0$ simulations once the system reaches a fully flattened quasi-steady state. 
This is shown in Fig. \ref{fig:Qflat_Q_vs_sS_beta0vs0001} for the case of $\beta=10^{-3}$. 
This contrasts with electrostatic simulations for which the turbulence cannot flatten the $q$ profile and the heat flux distinctly depends on the non-uniformity amplitude.
Note that finite $\beta$ has a small direct destabilizing effect, as can be seen by comparing the heat fluxes from the $\Tilde{s}_{1}^{S}=0$ simulations with $\beta = 0$ and $\beta = 10^{-3}$.

The simulations discussed so far have been performed without collisions. 
However, it is important to investigate the effects of collisions as the associated resistivity leads to current diffusion.
Collisions were modeled with the reduced form of the linearized Landau collision operator \cite{Rosenbluth1957, Abel2008}, including intra- and inter-species collisions, in simulations with $\hat{s}^{S}_{1}=0.006$.
Note that collisional diffusion of the current layers creating the imposed safety factor modulation is not modeled because the current is assumed to be carried by a small collisionless population of superthermal electrons driven by an ECCD source \cite{Ball2022}.
Our simulations revealed that the current diffusion from collisions causes the effective safety profile $\Bar{q}_{total}$ to partially relax back towards the imposed non-uniform profile $\tilde{q}$. 
The importance of this effect depends on collisionality.
For a normalized GENE collision frequency of $\nu = 0.005$ (equivalent to an electron-ion collision frequency of $\nu_{ei} = 0.38 v_{th,i}/R$), a stationary zonal component $ \langle A_{\parallel} \rangle_{y,z,t} \approx 0.8 \langle A_{\parallel, flat} \rangle_{y,z,t}$ is maintained, where $\langle A_{\parallel, flat} \rangle_{y,z,t}$ is the profile needed for full flattening.
However, at the higher frequency $\nu = 0.02$ (equivalent to an electron-ion collision frequency of  $\nu_{ei} = 1.52 v_{th, i}/R$), the zonal $\langle A_{\parallel} \rangle_{y,z,t}$ value decreased to $\approx 0.6 \langle A_{\parallel, flat} \rangle_{y,z,t}$.
We thus expect the safety factor flattening effect to be largest in the core of current fusion devices and to be even more prominent in future machines, which will have lower collisionality. 
This may also be related to the ITB heating power threshold \cite{Joffrin2002}. 
If we assume that turbulent self-interaction is an important mechanism in ITB triggering then reducing collisionality through the higher temperatures from heating would be beneficial as it would lower the resistive current diffusion.

So far in this section, we have only considered flattening about integer surfaces, but complete profile flattening was also observed in simulations with a finite parallel boundary shift $\Delta y_{0}$.
The shift was set to $\Delta y_{0} / L_{y}=0.25$ and $\Delta y_{0} / L_{y}=0.5$ to simulate $4^{th}$ and $2^{nd}$ order rational surfaces respectively.
The non-uniformity amplitude was fixed to $\hat{s}^{S}_{1}=0.004$ and collisions were not included.
This illustrates that safety factor profile flattening does not only occur at the lowest order integer surfaces. 
One can expect to see this effect on higher-order surfaces as well.

Overall, this subsection demonstrated that a system with a small but finite $\beta$, dominated by electrostatic instabilities, responds to a small imposed safety factor profile non-uniformity by generating turbulent currents that fully cancel the imposed non-uniformity, driving the $q$ profile toward rational values.
Consequently, the total heat flux increases to match the nominal case without the imposed non-uniformity. 
While collisions, through current diffusion, weaken this flattening effect, they do not eliminate it.

\subsection{Turbulent modification of safety factor profile with large non-uniformity and finite binormal shift}\label{sec:q_profile_scan}

In the previous section, we examined how turbulence reacts to imposed small amplitude modulations of a flat safety factor profile in weakly electromagnetic systems. 
In all cases, turbulent-generated currents flattened the non-uniformity. 
However, when the non-uniformity is substantial enough to introduce additional higher-order rational surfaces, the safety factor profile can flatten around multiple rational surfaces, resulting in a stepped safety factor profile, as in the linear magnetic shear simulations considered in Sec. \ref{sec:corrug_linear_profile}. 
Furthermore, by introducing a background binormal shift $\Delta y_{0}$ at the parallel boundary of the flux tube, we can scan the safety factor profile minimum for values close to but different from rational.
This approach allows us to emulate experimental conditions often associated with ITB formation and identify changes in turbulence characteristics as the minimum safety factor $q_{min}$ approaches and crosses a rational surface.

\begin{figure}[!hb]
\centering
\includegraphics[width=0.85\textwidth]{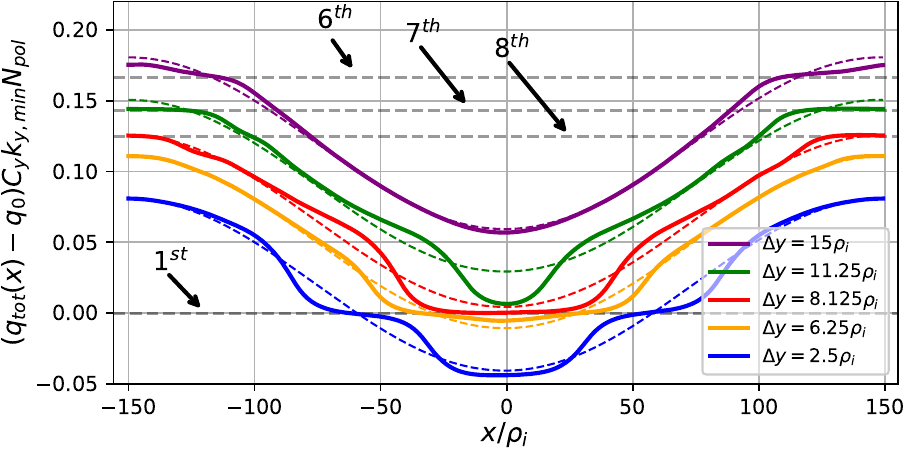}
\caption{A safety factor profile scan achieved by varying the background binormal shift $\Delta y_{0}$ for a weakly electromagnetic $\beta=10^{-4}$ system with a sinusoidal safety factor profile. The solid colored curves represent the turbulence-modified profiles, while the dashed colored curves indicate the initial imposed profiles. The turbulence-modified profiles were computed by averaging the final $10 \%$ of the simulation. The grey dashed horizontal lines denote the normalized safety factor values at various low order rational surfaces (1$^{st}$, 6$^{th}$, 7$^{th}$, and 8$^{th}$ order), as indicated in the plot. The simulations were performed using the parameters shown in Table \ref{tab:parameters_nonlinear_sS_etaScan_study} of \ref{app:simulation_parameters}.}
\label{fig:QFlat_sSm0025_etaScan_qProfile_v2}
\end{figure}

We conducted a safety factor profile scan by varying the background binormal shift $\Delta y_{0}$ for a case with $\hat{s}^{S}_{1}=0.025$, no background magnetic shear $\hat{s}_{0} = 0$, and $L_{x}=300 \rho_{i}$ (all parameters shown in Table \ref{tab:parameters_nonlinear_sS_etaScan_study}). 
Scanning $\Delta y_{0}$ modifies the offset $\Delta q$ in the safety factor profile as described in Eqs. \eqref{eq:delta_q_notnormalized} and \eqref{eq:total_imposed_q} and can be seen as corresponding to a scan in the safety factor profile minimum value $q_{min}$. 
For example, for considered parameters the sinusoidal modulation $\Tilde{q}(x)$ of the safety profile has an amplitude $\Delta \Tilde{q}$ such that $|\Delta \Tilde{q} C_{y} k_{y,min} N_{pol}| = 0.06$.
If we add a background shift of $\Delta y_{0} = 8.125 \rho_{i}$, this shifts the entire $q$ profile by $\Delta q C_{y}k_{y,min}N_{pol} \simeq 0.065$.
Adding these two contributions, a sinusoidal variation with an amplitude of $0.06$ and a constant offset of $0.065$, means that the total binormal shift at $x=0$ is just $(q_{min}-q_{0})C_{y}k_{y,min}N_{pol} \approx 0.005$.
In this case, the value of $q$ at $x=0$ is very close to, but not exactly integer.
The background binormal shift can thus be used to vary the minimum around the integer surface (either above or below).
To reduce computational costs and shorten the turbulent eddy parallel length to avoid significant parallel self-interaction at even high-order rational surfaces, we used artificially heavy electrons, i.e. $m_{i}/m_{e}=368$.  
Plasma $\beta$ was set to $\beta=10^{-4}$ to ensure that electromagnetic microinstabilities remained subdominant. 
We performed a safety factor scan beginning with a profile significantly above the integer surface, lowering it to the integer surface, and then lowering it further to include the integer surfaces at two significantly separated radial locations. 
The initial and turbulence-modified profiles from the scan are shown in Fig. \ref{fig:QFlat_sSm0025_etaScan_qProfile_v2}.

Several features of the modified safety factor profile are noteworthy. 
First, the profile flattens around low-order rational surfaces, evident at the 1$^{st}$, 6$^{th}$, 7$^{th}$, and 8$^{th}$ order rational surfaces. 
Second, even when the initial profile does not cross a rational surface, the modified profile is ``pulled" towards the nearest lowest order rational surface, as seen in the cases with $\Delta y_{0} = 8.125 \rho_{i}$ and $\Delta y_{0} = 11.25 \rho_{i}$. 
This illustrates that, as the safety factor profile minimum $q_{min}$ approaches a low-order rational value, turbulent-driven zonal currents can locally modify the safety factor profile and pull it towards the rational value, potentially aiding in triggering an ITB.

\begin{figure}[!hb]
\centering
\includegraphics[width=0.85\textwidth]{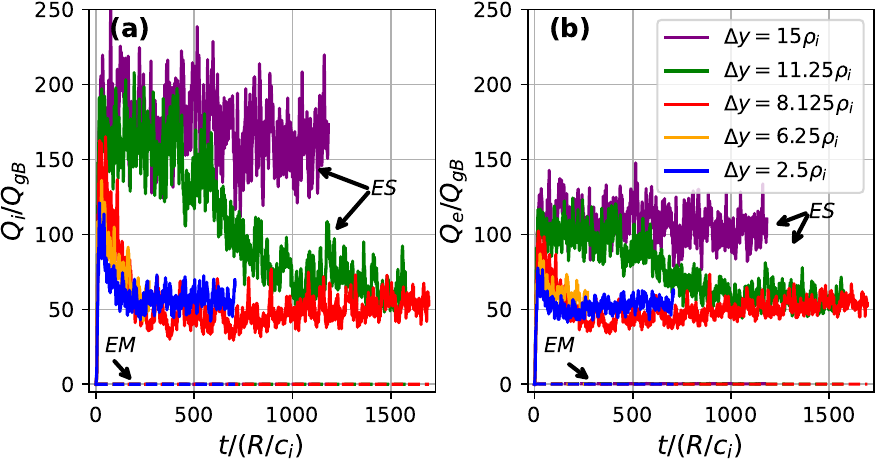}
\caption{Time traces of (a) ion and (b) electron electrostatic and electromagnetic heat fluxes for CBC drive at fixed plasma $\beta=10^{-4}$, $\hat{s}_{0}=0$ and $\Tilde{s}^{S}_{1} = 0.025$ with varying background binormal shift $\Delta y_{0}$. The corresponding safety factor profiles are shown in Fig. \ref{fig:QFlat_sSm0025_etaScan_qProfile_v2}. The simulations were performed using the parameters shown in Table \ref{tab:parameters_nonlinear_sS_etaScan_study} of \ref{app:simulation_parameters}.}
\label{fig:QFlat_sSm0025_s0_Qtrace_etaScan}
\end{figure}

The heat flux time traces for these simulations are shown in Fig. \ref{fig:QFlat_sSm0025_s0_Qtrace_etaScan}. 
The three simulations with the initial profiles close to or crossing the lowest order rational surface have about three times lower total heat flux than the $\Delta y_{0} =15 \rho_{i}$ simulation, which is not close to any low-order rational surfaces.
It is interesting to observe that the ion channel is stabilized more than the electron channel. 
To match the same heat fluxes between the $\Delta y_{0} = 8.125 \rho_{i}$ and $\Delta y = 15 \rho_{i}$ cases, all the logarithmic gradients (i.e. density, ion temperature, and electron temperature gradients) for the $\Delta y_{0} = 15 \rho_{i}$ case had to be reduced to $\sim 75 \%$ of their initial values. 
This indicates that in a flux-driven scenario, the background gradient for the $\Delta y_{0} = 8.125 \rho_{i}$ case would have to steepen to maintain a constant heat flux, potentially representing the formation of a transport barrier.
Finally, it is noteworthy that the $\Delta y_{0} = 11.25 \rho_{i}$ simulation initially exhibits high heat flux, which progressively decreases as turbulence-generated currents slowly build up and the safety factor profile minimum finally evolves toward the lowest order rational surface.

Overall, these flux tube simulations exhibit a reduction in transport as the reversed shear safety factor profile crosses a low-order rational surface.
This is consistent with the experimental observations of ITB triggering and further demonstrates the role of turbulent currents in modifying the safety factor profile near low-order rational surfaces.

\subsection{Turbulent modification of pseudo-global and global reversed shear safety factor profiles}\label{sec:global_sims}
In this section, we present first-of-a-kind pseudo-global simulations based on the flux tube version of GENE code, which combines a linear safety factor profile with $\hat{s}_{0} \neq 0$ with a finite sinusoidal modulation $\Tilde{q}(x)$ as given by Eqs. \eqref{eq:total_imposed_q} and \eqref{eq:non_uniform_q}.
These simulations are designed to model reversed magnetic shear $q$ profiles as accurately as possible.
Such a safety factor profile features a minimum at some radial location, which is favorable for triggering an ITB.
In the framework of our modified flux tube model, this setup thus allows us to consider the radial safety factor profile most relevant for ITB formation.
We also compare these results with global gyrokinetic ORB5 simulations that feature a reversed magnetic shear profile.
A more detailed discussion of the results obtained with ORB5 is provided in a companion paper \cite{DiGiannatale2024_GlobalReversedShear}.
In Ref. \cite{DiGiannatale2024_GlobalReversedShear}, it is clearly demonstrated that the flattening of the safety factor profile, due to turbulent self-generated currents around rational surfaces, is not an artifact of flux tube simulations, but is also observed in standard global simulations.
This provides a strong validation of our approach and the results discussed in this paper.

\begin{figure}[!hb]
\centering
\includegraphics[width=0.65\textwidth]{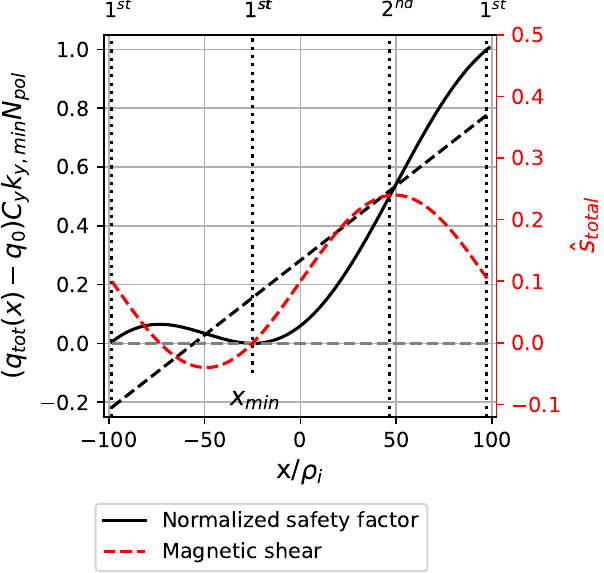}
\caption{The total imposed non-uniform safety factor profile (black solid line), the background linear part of the safety factor profile (black dashed line), and the total background magnetic shear profile $\hat{s}_{tot}(x)$ (red dashed line), plotted as functions of the radial position. These profiles were generated using a single Fourier sine coefficient,  $\hat{s}^{S}_{1} \approx 0.14$, with an average background magnetic shear of $\hat{s}_{0} = 0.1$, and $q_{min}$ position at $x_{min} \approx -25 \rho_{i} $. Black vertical dotted lines mark the locations of first and second-order rational surfaces.
}
\label{fig:QFlat_s01_sSm014_example_140824}
\end{figure}

The safety factor profile used in our pseudo-global simulations is shown in Fig. \ref{fig:QFlat_s01_sSm014_example_140824}, with $q_{min}$ coinciding with an integer surface.
We also consider a simulation where the minimum $q$ value is shifted away from this integer surface by applying a background binormal shift, $\Delta y_{0} \neq 0$, without altering the shear profile.
As in standard flux tube simulations, all other radial plasma profiles are initially linear, meaning they have constant gradients.
The initial plasma gradient values chosen for these simulations are those for CBC.
As the nonlinear simulations evolve, stationary zonal components of the fluctuating quantities develop and lead to periodic modulations of these gradient profiles, in a similar way as illustrated in Fig. \ref{fig:QFlat_s01_b0vsb0001_ProfileCorrugations_300524} for standard flux tube simulations.
It is important to note that the radial boundary condition remains periodic.
This implies that, regardless of the strong plasma profile corrugations, the domain-averaged gradients of these profiles remain fixed.
Additionally, the background values of density, temperature, safety factor, etc. are still assumed to be constant across the radial domain when calculating quantities in the gyrokinetic equation.
These approximations become less accurate when large, non-uniform safety factor profile shearing effects are included to model reversed shear regions that span several tens of $\rho_{i}$.
In this case, certain $\rho^{*}$ effects related to safety factor profile shearing are introduced, while the other plasma profiles (e.g., density, temperature, magnetic geometry coefficients) are treated in the local limit.
Nonetheless, as we will discuss shortly, the agreement with global ORB5 results suggests that while flux tube pseudo-global simulations are not appropriate for modeling a large ITB, they are able to capture key physical phenomena related to the safety factor profile shape with minimal computational cost.

\begin{figure}[!hb]
\centering
\includegraphics[width=0.85\textwidth]{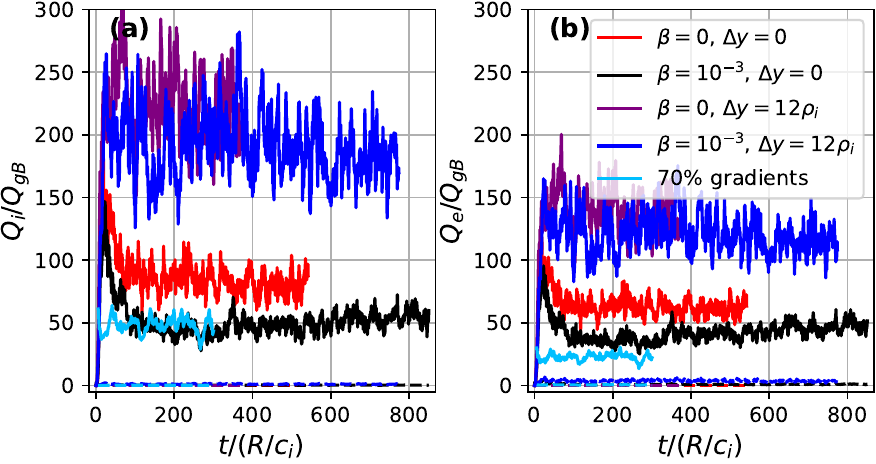}
\caption{Time traces of the (a) ion and (b) electron electrostatic and electromagnetic heat fluxes with $\hat{s}_{0}=0.1$ and $\hat{s}^{S}_{1} = -0.14$ with varying $\beta$, $q_{min}$ values, and density and temperature gradients. The simulations were performed using the parameters shown in Table \ref{tab:parameters_nonlinear_pseudoglobal_study} of \ref{app:simulation_parameters}.}
\label{fig:Qtrace_CBC_b0vsb0001_s01_sSm014_eta28_120824}
\end{figure}

To begin with, it is illustrative to compare the average total ion and electron heat flux from simulations using a purely linear safety factor profile with $\hat{s}_{0}=0.1$ and from simulations with an added sinusoidal modulation $\Tilde{q}(x)$ leading to a reversed shear profile (shown in Fig. \ref{fig:QFlat_s01_sSm014_example_140824}).
We will first compare electrostatic results, for which the safety factor profile is not modified by the turbulence.
As shown in Figs. \ref{fig:QFlat_s01_Qtrace_betaScan_080524} and \ref{fig:QFlat_s01_mex50_Qtrace_betaScan_080524}, purely linear $q$ simulation with the normal electron mass $m_{i}/m_{e} = 3680$ have $Q_{tot}/Q_{gB} \approx 210 $, while using heavier electrons with mass of $m_{i}/m_{e} = 184$ gives $Q_{tot}/Q_{gB} \approx 180 $.
Here $Q_{tot} = Q_{i}+Q_{e}$ is the sum of average ion and electron heat fluxes.
In comparison, the electrostatic reversed shear simulation studied in this section, which has heavy electrons $m_{i}/m_{e} = 368$ and the same average background magnetic shear $\hat{s}_{0} = 0.1$, shows a much lower average total heat flux of $Q_{tot}/Q_{gB} \approx 100 $ (see red line in Fig. \ref{fig:Qtrace_CBC_b0vsb0001_s01_sSm014_eta28_120824}).
This demonstrates that, while the average magnetic shear $\hat{s}_{0} = 0.1$ is the same, the variation in the safety factor profile and the presence of a low magnetic shear region coinciding with an integer safety factor value, reduces the heat flux by approximately $\sim 50 \%$.

In Fig. \ref{fig:Qtrace_CBC_b0vsb0001_s01_sSm014_eta28_120824}, we compare the heat flux time traces of five pseudo-global simulations with a reversed shear safety factor profile.
The simulations corresponding to the black and red time traces in Fig. \ref{fig:Qtrace_CBC_b0vsb0001_s01_sSm014_eta28_120824} both have a safety factor minimum coinciding with integer surfaces as shown in Fig. \ref{fig:QFlat_s01_sSm014_example_140824}.
The difference between these simulations is only the value of $\beta$, with the red case being electrostatic and the black case having $\beta = 10^{-3}$.
We see that adding electromagnetic effects reduces the heat flux.
However, the heat flux is changed more dramatically by shifting the safety factor profile minimum away from the integer surface (by setting $\Delta y_{0} = 12 \rho_{i}$ or equivalently $(q_{min}-q_{0})C_{y}k_{y,min}N_{pol} \approx 0.095$).
These simulations exhibit more than three times larger total heat flux regardless of the value of $\beta$ (comparing either blue or purple time traces with the red in Fig. \ref{fig:Qtrace_CBC_b0vsb0001_s01_sSm014_eta28_120824}).
This is consistent with the results presented in Fig. \ref{fig:QFlat_sSm0025_s0_Qtrace_etaScan}.
Overall, Fig. \ref{fig:Qtrace_CBC_b0vsb0001_s01_sSm014_eta28_120824} demonstrates that, while the main stabilization arises from strong parallel self-interaction due to $q_{min}$ coinciding with an integer value, there is a further stabilizing effect when one allows electromagnetic effects to flatten the $q$ profile around the integer value.

\begin{figure}[!hb]
\centering
\includegraphics[width=0.55\textwidth]{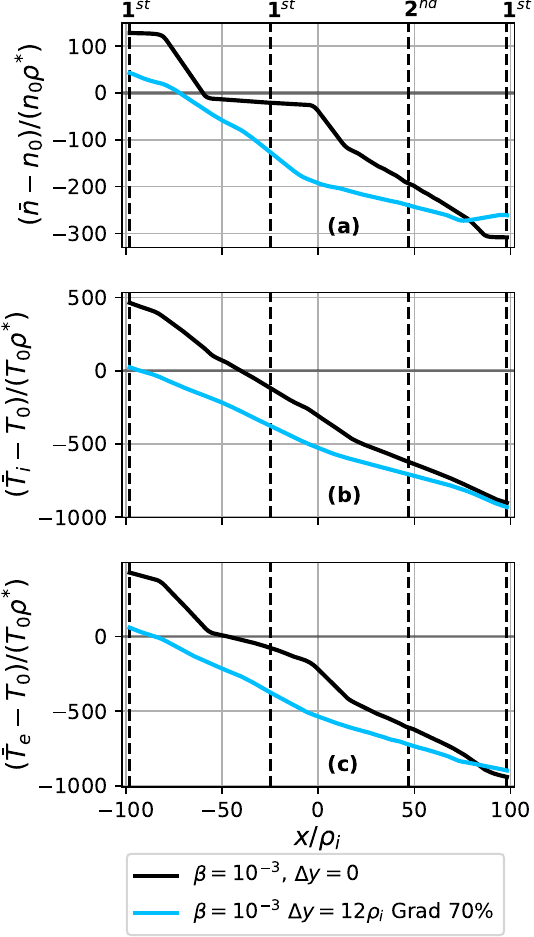}
\caption{Radial profiles of the flux surface and time-averaged (a) density, (b) ion temperature, and (c) electron temperature. The black solid line represents simulation with $q_{min}$ at an integer surface and CBC gradients. The cyan solid line corresponds to a simulation away from the integer surface, with gradients reduced to 70$\%$ of the CBC values in order that both simulations have the same heat flux. The black vertical dotted lines indicate integer surfaces up to the 2nd order. The simulations were performed using the parameters shown in Table \ref{tab:parameters_nonlinear_pseudoglobal_study} of \ref{app:simulation_parameters}.}
\label{fig:QFlat_sSm014_etaScan_plasmaProfile_nomVS70pGrad_v2_241024}
\end{figure}

It is important to highlight that the simulations discussed so far were gradient-driven, i.e. their fluxes differ while their radially averaged gradients remain fixed.
To provide a more practical measure of the stabilizing effect of an integer $q_{min}$, we again performed a flux-matching exercise where the gradients for the blue case shown in Fig. \ref{fig:Qtrace_CBC_b0vsb0001_s01_sSm014_eta28_120824} were reduced until the heat flux matched that of the black case.
A 30$\%$ reduction in the gradients was necessary to approximately match the heat fluxes, i.e. setting the normalized density gradients to $\omega_{n}=1.55$ and the temperature gradients to $\omega_{T_{i}}=\omega_{T_{e}}=4.87$.
The result is represented by the cyan time trace in Fig. \ref{fig:Qtrace_CBC_b0vsb0001_s01_sSm014_eta28_120824}.
The comparison between the total, turbulence-modified temperature and density profiles of the black and cyan simulations is shown in Fig. \ref{fig:QFlat_sSm014_etaScan_plasmaProfile_nomVS70pGrad_v2_241024}.
For the simulations with $q_{min}$ corresponding to a 1$^{st}$ order rational surface, a rudimentary transport barrier can be identified in all channels, and it is most pronounced in the ion temperature profile.
The density and electron temperature profiles are less significantly affected due to profile flattening at the rational surface.

\begin{figure}[!hb]
\centering
\includegraphics[width=0.55\textwidth]{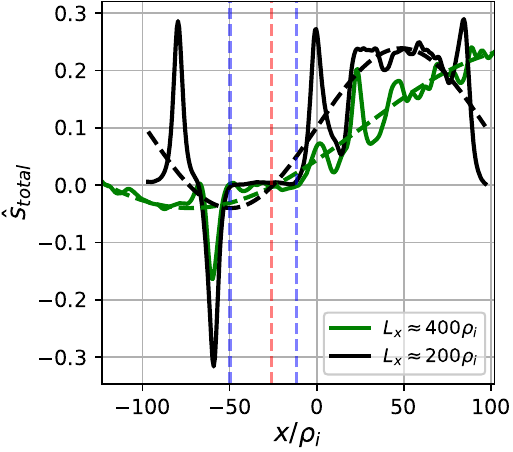}
\caption{Radial profiles of the total magnetic shear for two simulations with different radial and binormal domain sizes. The green solid and broken lines represent the modified and imposed shear profiles, respectively, for the simulation with a minimal binormal wave number of $k_{y,\text{min}} \rho_{i} = 0.025$ and radial domain size $L_{x} = 400 \rho_{i}$. The black solid and broken lines show the modified and imposed shear profiles for the simulation with $k_{y,\text{min}} \rho_{i} = 0.05$ and $L_{x} = 200 \rho_{i}$. The dashed vertical blue lines mark the boundaries of the zero magnetic shear region, and the red broken line indicates the initial radial location of $q_{min}$. The simulations were performed using parameters shown in Table \ref{tab:parameters_nonlinear_pseudoglobal_study} of \ref{app:simulation_parameters}, with the green simulation having twice as many radial and binormal grid points.}
\label{fig:QFlat_CBC_b0vsb0001_s01_sSm014_eta028_plasmaProfile_120824}
\end{figure}

Finally, we compare electromagnetic simulations with $q_{min}$ coinciding with the lowest order rational surface for different radial widths of the flux tube, specifically $L_{x} \approx 200 \rho_{i}$ and $L_{x} \approx 400 \rho_{i}$, but at fixed values of $\hat{s}_{0}=0.1$ and $\Tilde{s}^{S}_{1}=0.14$.
Varying $L_{x}$ thus modifies the curvature of the safety factor profile according to Eq. \eqref{eq:non_uniform_q}.
We find that the width of the flattened region around integer $q$ values does not depend on $L_{x}$ and remains fixed in units of ion gyro radius.
The turbulence-modified magnetic shear profiles are shown in Fig. \ref{fig:QFlat_CBC_b0vsb0001_s01_sSm014_eta028_plasmaProfile_120824}.
We see a $\sim 35 \rho_{i}$ wide near-zero magnetic shear ($\hat{s} \approx 0$) region localized at the central integer surface, i.e. where the imposed safety factor profile $q_{tot}$ has its minimum.
This region is asymmetric about the initial radial location of the integer surface due to the larger magnetic shear and higher density of low order rational surfaces present to its right.
Overall, the invariant width of the zero-shear region in units of $\rho_{i}$ suggests that the observed flattening of the safety factor around integer $q_{min}$ is a turbulence-scale ($\sim \rho_{i}$) phenomenon in the limit of small $\rho_{*} \ll 1$.

In a companion paper \cite{DiGiannatale2024_GlobalReversedShear}, similar overall trends are observed.
In particular, global ORB5 simulations of reversed shear profiles exhibit parallel velocity corrugations that generate parallel currents and modify the safety factor profile around rational $q$ surfaces. 
Flattening is observed around the minimum of the safety factor profile when $q_{min}$ is an integer in both gradient-driven and flux-driven ORB5 simulations.
Additionally, a comparison of transport between simulations with $q_\text{min} = 2$ and those with $q_\text{min}$ away from an integer value confirms the globally stabilizing effect of $q_\text{min} = 2$, due to a complex mechanism involving strong self-interaction at that surface.
Additionally, in electromagnetic flux-driven simulations initialized with $q_{min} = 2.01$, turbulent-generated currents ``pull" the safety factor profile minimum towards the integer value $q_{min} = 2$ surface.
This behavior is identical to the flux tube simulation shown in the green curve of Fig. \ref{fig:QFlat_sSm0025_etaScan_qProfile_v2}. 
Similarly, ORB5 finds that this change in the safety factor profile significantly reduces transport further highlighting the stabilizing effects of an integer $q_{min}$ and the turbulence-induced modifications to the safety factor profile.
These observations further support the central idea of this paper: turbulence-generated currents lead to a stepped safety factor profile, which in turn significantly reduces turbulent transport. 
The agreement between flux tube and global simulations validates the physical relevance of these effects and indicates that they may play an important role in ITB triggering.

Besides the fundamentally different numerical approaches of the two codes (Eulerian for GENE and PIC for ORB5), it is also worth noting that the agreement was achieved despite additional differences in the physical conditions considered in the simulations. 
The global ORB5 simulations investigated a turbulent regime with different driving gradients compared to the flux tube GENE simulations, featuring finite ion and electron temperature gradients based on CBC values, but without a density gradient.
This setup allowed for ITG and temperature-gradient-driven TEM turbulence but did not support density-gradient-driven TEM modes.
In contrast, most of the GENE simulations presented in this paper were conducted with CBC gradients that included both temperature and density gradients.
In some limited cases of interest (see Figs. \ref{fig:QFlat_s01_Apar_betaScan_070524}, \ref{fig:Qflat_Q_vs_sS_pITGandCBC_300424} and \ref{fig:QFlat_small_sSm_pITGandCBC_upar_sSmScan_300424}), comparisons were also made with pITG simulations, where only the ion temperature gradient was non-zero.
Despite these three distinct setups the primary results concerning turbulent modifications to the safety factor profile and stabilization of turbulence remained consistent across all simulations.

Lastly, zonal flows were found to play a role in the transport reduction in both global and flux tube simulations.
In global simulations, however, radially extended eddies (as measured in terms of $\phi$ correlation) were found to mediate transport when $q_\text{min}$ is sufficiently far from a rational surface. 
In contrast, when $q_\text{min}$ coincides with a rational surface, these radially extended eddies were strongly cut by zonal flows generated in the vicinity of $q_\text{min}$.
While zonal shear flows also form in the flux tube simulations studied in this paper, their strength and impact on radial eddy size and transport seem to be weaker, aligning with previous flux tube work on the stabilizing role of turbulent self-interaction \cite{Ajay2020}.
This could be a result of the different radial eddy structure observed in ORB5, though why it does not appear the same in GENE remains unclear.
Nonetheless, both global and flux tube simulations find that steady zonal flows arise when $q_\text{min}$ is an integer even when this is a result of turbulent modifications to the safety factor profile.
The observed differences in turbulence and zonal flow strength between global and flux tube simulations may stem from differences in the turbulent regimes or from the nature of global versus flux tube simulations.
This warrants further investigation, particularly to understand the role of system size, boundary conditions, and safety factor profiles in shaping the zonal flow behavior.

Overall, considering the different physical parameters and numerical techniques employed between global and flux tube simulations, the observed agreement of turbulent self-generated currents modifying the safety factor profile around rational $q$ values in reversed shear configurations is remarkable, giving greater confidence in the physical validity of these results.

\section{Conclusions}

In this paper, we have investigated the impact of ion scale turbulence-generated current layers on the safety factor profile.
The parameter regime we investigated, involving low magnetic shear in proximity to low-order rational surfaces, was chosen to be relevant to ITB formation.
We demonstrated that turbulence intensity inhomogeneity close to low-order rational surfaces can drive turbulent currents that lead to significant stationary corrugations in the safety factor profile.
This produces a stepped safety factor profile and wide zero magnetic shear regions around these rational surfaces.
The modified safety factor profile acts back on turbulence, leading, in most extreme cases, to a four-fold reduction in the heat flux (see Fig. \ref{fig:QFlat_s01_Qtrace_betaScan_080524}) and significant changes in the gradients of density and temperature that resemble a transport barrier.

We have performed the study with three different approaches: (i) standard local GENE simulations with constant low but finite magnetic shear, (ii) pseudo-global GENE simulations with non-uniform magnetic shear, and (iii) global ORB5 simulations with reversed shear safety factor profiles (discussed in greater detail in companion paper \cite{DiGiannatale2024_GlobalReversedShear}).
All of these simulations consistently exhibit the flattening of externally imposed safety factor profiles at rational surfaces by turbulent currents and a strong subsequent impact on turbulent transport when magnetic shear is low.
This suggests that our results are a robust feature in tokamaks and provide a possible triggering mechanism for ITB formation.

While several questions have been addressed and important conclusions drawn, many avenues for future research remain.
We have demonstrated the importance of turbulence-generated current under plasma core conditions relevant to ITB formation.
It is intriguing to consider which, if any, aspects of the physical mechanisms discussed here could also apply to edge transport barriers.
It is not clear if the larger relative turbulence amplitudes (and stronger associated currents) in the edge region would be enough to overcome generally much larger magnetic shear.
However, in certain scenarios, this limitation might not apply, as equilibrium reconstructions suggest that the bootstrap current can produce radially localized low magnetic shear regions in the plasma edge \cite{Kessel2007_AdvancedOP_ITER, Zheng2013_safety_factor_plateau, Igochine2017_teringmodecrash}.
Given these considerations, the interplay between turbulent self-interaction, turbulent-generated currents and the safety factor profile in the plasma edge warrants a more detailed investigation.

Additionally, we have limited the scope of our study to ion-scale turbulence.
We have \textit{a priori} assumed that electron scale turbulence does not play a significant role in our simulations. 
However, further research is needed to understand the potential impact of electron-scale turbulence on safety factor profile flattening as well as its response to changes in the effective safety factor profile due to ion-scale turbulence. Furthermore, $\beta$ was limited to low values of $\beta \leq 10^{-3}$ to remain dominated by electrostatic instabilities. An investigation into the impact of fully electromagnetic turbulence on the safety factor profile around rational values is needed as it may be even more important. Finally, while our study was conducted within the context of ITB physics in tokamaks, the results are highly relevant to stellarators as well, where the global magnetic shear is often low. 

\ack
The authors would like to thank Oleg Krutkin, Alessandro Geraldini, Ben McMillan, and Antoine Hoffmann for useful discussions pertaining to this work.
This work has been carried out within the framework of the EUROfusion Consortium, via the Euratom Research and Training Programme (Grant Agreement No 101052200 — EUROfusion) and funded by the Swiss State Secretariat for Education, Research and Innovation (SERI). Views and opinions expressed are however those of the author(s) only and do not necessarily reflect those of the European Union, the European Commission, or SERI. Neither the European Union nor the European Commission nor SERI can be held responsible for them.
We acknowledge the CINECA award under the ISCRA initiative, for the availability of high performance computing resources and support. 
This work was supported by a grant from the Swiss National Supercomputing Centre (CSCS) under project ID s1050 and s1097. 
This work was supported in part by the Swiss National Science Foundation.

\appendix

\section{Turbulent current drive via electron momentum flux}\label{app:turb_current_drive}
Here we outline a derivation of Eq. \eqref{eq:current_eq_full}, the evolution equation for the parallel electron current, from gyrokinetics and use it to identify the main physical mechanism driving the current.
We will omit electromagnetic effects in this derivation as they were found to be insignificant for the current drive when $\beta \leq 10^{-3}$ and consider the electrostatic version of Eq. \eqref{eq:simple_GK_eq}.
Our notation and part of the derivation closely follow reference \cite{Lapillonne2010Thesis}.

Since $m_{e} \ll m_{i}$, we assume that electron current dominates the total current (which we find to be true in our numerical simulations), implying the total perturbed current is
\begin{equation}
    \label{eq:current_electron}
    j_{1,\parallel} \simeq j_{1,\parallel,e} =  C_{\mu_{0}} \int B_{0}(z) v_{\parallel} \Bar{f}_{1,e}(\mathbf{x}, v_{\parallel}, \mu) dv_{\parallel} d \mu = \langle  \Bar{f}_{1,e}  \rangle_{10},
\end{equation}
where $\langle ... \rangle_{10}$ denotes the first parallel velocity moment.
Note that, for convenience, we collect the constants in front into a single constant $C_{\mu_{0}} = 2\pi \mu_{0} q_{e}/m_{e}$.

Next, we will consider the local limit and use the field-aligned coordinate system $(x,y,z)$. 
In these coordinates the equilibrium magnetic field can be written as
\begin{equation}
    \label{eq:magnetic_field}
    \mathbf{B}_{0} = C_{xy} \nabla x \times \nabla y,
\end{equation}
where $C_{xy}$ is a normalization factor. 
In the local limit, $C_{xy}$ is a constant and the equilibrium quantities and their gradients depend only on the poloidal coordinate $z = \chi$. 

Using these assumptions, we write out the electrostatic version of Eq. \eqref{eq:simple_GK_eq} explicitly in the field-aligned coordinate system to arrive at (similar to Eq. (2.62) in Ref. \cite{Lapillonne2010Thesis})
\begin{align}
    \label{eq:GK_inFieldAlignedCoord}
    \partial_{t} \Bar{f}_{1,e} & + \frac{1}{C_{xy} \gamma_{1}} [ \gamma_{2}\partial_{z} f_{0,e} \partial_{x} \Bar{\phi}_{1} - \gamma_{1} \partial_{x} f_{0,e} \partial_{y} \Bar{\phi}_{1}+\gamma_{3} \partial_{z} f_{0,e} \partial_{y} \Bar{\phi}_{1}   \notag \\
    & - \frac{\mu}{m_{e} v_{\parallel}} (\gamma_{2} \partial_{z} B_{0} \partial_{x} \Bar{\phi}_{1} - \gamma_{1} \partial_{x} B_{0} \partial_{y} \Bar{\phi}_{1} + \gamma_{3} \partial_{z} B_{0} \partial_{y} \Bar{\phi}_{1}) \partial_{v_{\parallel}} f_{0,e} ] \notag \\
    & + \frac{1}{C_{xy}} (\partial_{x} \Bar{\phi}_{1} \Gamma_{y,e}-\partial_{y} \Bar{\phi}_{1} \Gamma_{x,e}) \notag \\
    & + \frac{\mu B_{0}+ m_{e} v_{\parallel}^{2}}{m_{e} \Omega_{e}} (\mathcal{K}_{x} \Gamma_{x,e}+\mathcal{K}_{y} \Gamma_{y,e}) + \frac{1}{C_{xy}} \frac{\mu_{0} v_{\parallel}^{2}}{\Omega_{e} B_{0}} \partial_{x} p_{0} \Gamma_{y,e} \notag \\
    & + \frac{C_{xy} v_{\parallel}}{ B_{0} J_{xyz}} \Gamma_{z,e} - \frac{C_{xy} \mu}{m_{e} B_{0} J_{xyz}} \partial_{z} B_{0} \partial_{v_{\parallel}} \Bar{f}_{1,e} = 0,
\end{align}
where $p_{0}$ is the background pressure,
\begin{equation}
    \Gamma_{i,e} = \partial_{i} \Bar{f}_{1,e} - \frac{q_{e}}{m_{e} v_{\parallel}} \partial_{i} \Bar{\phi}_{1} \frac{\partial f_{0,e}}{\partial v_{\parallel}}, \qquad i\in \{x,y,z\},
\end{equation}
$\gamma_{i}$ is a combination of geometric coefficients
\begin{align}
    & \gamma_{1} = g^{xx}g^{yy}-g^{xy}g^{xy}, \\
    & \gamma_{2} = g^{xx}g^{yz}-g^{xy}g^{xz}, \\
    & \gamma_{3} = g^{xy}g^{yz}-g^{yy}g^{xz},
\end{align}
and
\begin{align}
    & \mathcal{K}_{x} = - \frac{1}{C_{xy}} \frac{\gamma_{2}}{\gamma_{1}} \partial_{z} B_{0}, \\
    & \mathcal{K}_{y} = \frac{1}{C_{xy}}(\partial_{x} B_{0} - \frac{\gamma_{3}}{\gamma_{1}} \partial_{z} B_{0}),
\end{align}
are additional geometric factors.

To further simplify Eq. \eqref{eq:GK_inFieldAlignedCoord}, we consider the background distribution function to be a local Maxwellian
\begin{equation}
    \label{eq:local_Maxwellian}
    f_{0,e}(z,v_{\parallel}, \mu) = \left(\frac{m_{e}}{2 \pi T_{0,e}}\right)^{\frac{3}{2}} n_{0,e} \exp{\left(-\frac{m_{e}v_{\parallel}^{2}/2+\mu B_{0}(z)}{T_{0,e}}\right)}
\end{equation}
as is standard in the gyrokinetic ordering.
This allows us to eliminate a number of terms related to background distribution from Eq. \eqref{eq:GK_inFieldAlignedCoord} and arrive at
\begin{align}
    \label{eq:GK_inFieldAlignedCoord_simple}
    \partial_{t} \Bar{f}_{1,e} & = \frac{1}{C_{xy}} \left[ \frac{d \ln{(n_{0,e})}}{dx} + \left(\frac{m_{e}v_{\parallel}^{2}}{2T_{0,e}}+\frac{\mu B_{0}}{T_{e}} - \frac{3}{2}\right) \frac{d \ln{(T_{0,e})}}{dx}\right] \partial_{y} \Bar{\phi}_{1}\notag \\
    & - \frac{1}{C_{xy}} (\partial_{x} \Bar{\phi}_{1} \Gamma_{y,e}-\partial_{y} \Bar{\phi}_{1} \Gamma_{x,e}) \notag \\
    & - \frac{\mu B_{0}+ m_{e} v_{\parallel}^{2}}{m_{e} \Omega_{e}} (\mathcal{K}_{x} \Gamma_{x,e}+\mathcal{K}_{y} \Gamma_{y,e}) - \frac{1}{C_{xy}} \frac{\mu_{0} v_{\parallel}^{2}}{\Omega_{e} B_{0}} \partial_{x} p_{0} \Gamma_{y,e} \notag \\
    & - \frac{C_{xy} v_{\parallel}}{ B_{0} J_{xyz}} \Gamma_{z,e} + \frac{C_{xy} \mu}{m_{e} B_{0} J_{xyz}} \partial_{z} B_{0} \partial_{v_{\parallel}} \Bar{f}_{1,e}.
\end{align}

After taking the first parallel velocity moment, performing a flux surface average, and neglecting the electron FLR terms (i.e. $\Bar{\phi}_{1} \approx \phi_{1}$), we arrive at an evolution equation for the zonal parallel current evolution
\begin{align}
\label{eq:current_eq_full_final_appendix}
     \partial_{t} \langle j_{1, \parallel} \rangle_{y,z} = & \underbrace{\frac{q_{e}}{C_{xy}}\partial_{x} \langle  \partial_{y} \phi_{1} u_{1, \parallel,e} \rangle_{y,z}}_{\textbf{(1)}} \notag \\
     & - \underbrace{\partial_{x} \left\langle \frac{\mathcal{K}_{x}}{2\pi\Omega_{e}} C_{\mu_{0}}  (q_{1, \perp,e} + 2 q_{1,\parallel,e} +4 p_{0} u_{1,\parallel,e}) \right\rangle_{y,z}}_{\textbf{(2)}} \notag \\
     & - \underbrace{\left\langle \frac{C_{\mu_{0}} C_{xy}}{J_{xyz} B_{0}} \frac{m_{e}}{4 \pi} \partial_{z}\left( T_{1,\parallel, e} n_{0,e} + T_{0,e}n_{1,e} \right) \right\rangle_{y,z}}_{\textbf{(3)}} \notag \\
     & + \underbrace{\left\langle \frac{C_{\mu_{0}} C_{xy}}{J_{xyz} B_{0}}  \frac{q_{e} \partial_{z} \phi_{1}}{2 \pi}  n_{0e}  \right\rangle_{y,z}}_{\textbf{(4)}} \notag \\
     &  + \underbrace{\left\langle \frac{C_{\mu_{0}} C_{xy} }{J_{xyz} B_{0}} \frac{\partial_{z} B_{0}}{B_{0}} \frac{m_{e}}{4 \pi} \left( T_{1,\parallel, e} n_{0,e} - T_{1, \perp,e } n_{0,e} \right) \right\rangle_{y,z}}_{\textbf{(5)}},
\end{align}
where $q_{1,\perp,e}$ is the perpendicular and $q_{1,\perp,e}$ is the parallel component of the parallel electron heat current density.
In this equation,
\begin{itemize}
    \item term \textbf{(1)} arises from the nonlinear term in the gyrokinetic equation and represents the divergence of the parallel electron momentum flux,
    \item term \textbf{(2)} arises from the toroidal curvature terms in the gyrokinetic equation and corresponds to the divergence of the parallel heat currents,
    \item terms \textbf{(3)} and \textbf{(4)} arise from the parallel advection term and correspond to the acceleration due to parallel pressure anisotropy and parallel electric field, 
    \item term \textbf{(5)} arises from a combination of the parallel advection and mirror term.
\end{itemize}

\begin{figure}[!hb]
\centering
\includegraphics[width=1\textwidth]{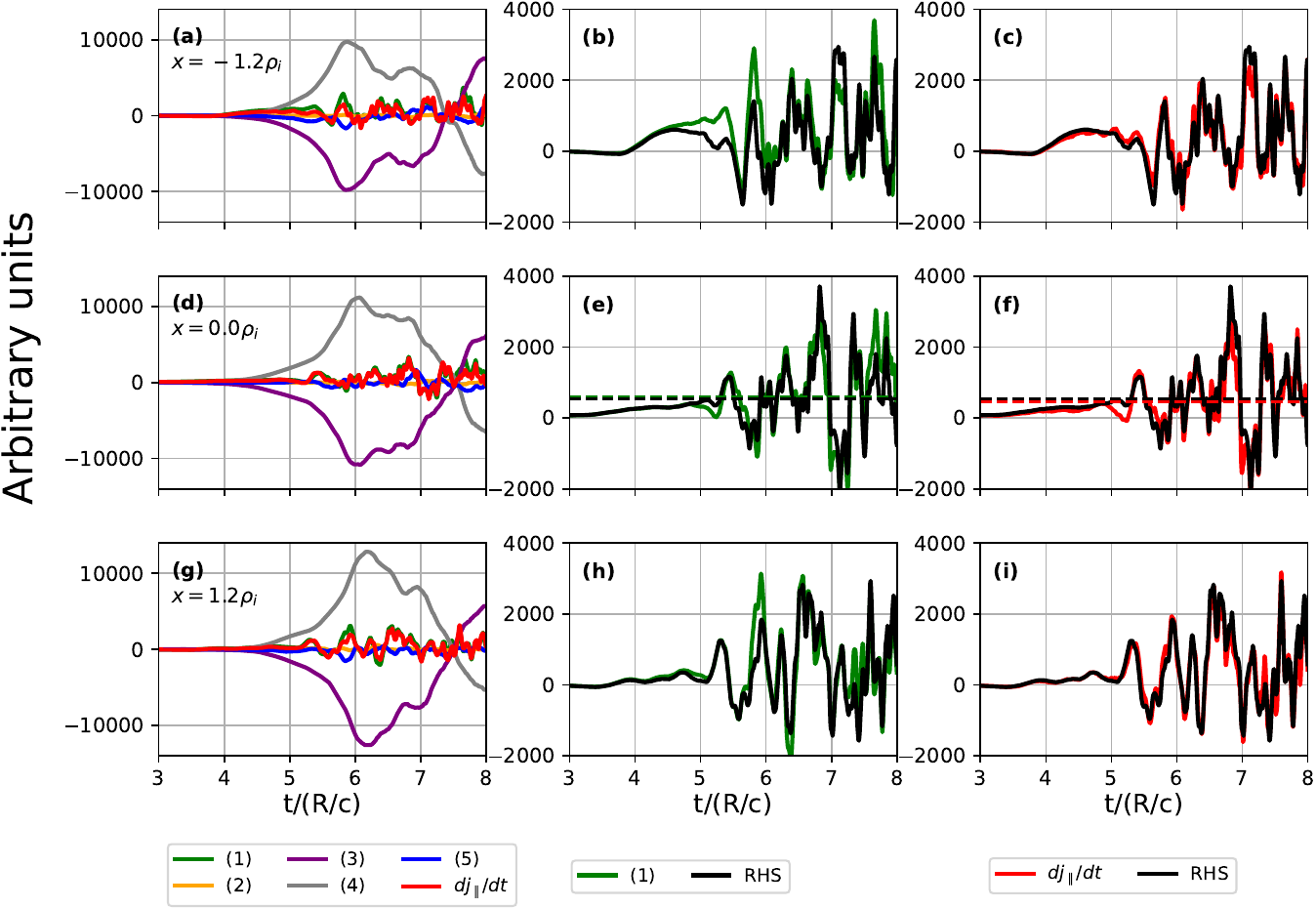}
\caption{(a, d, g) Time traces of different terms in Eq. \eqref{eq:current_eq_full_final_appendix}. (b, e, h) A comparison of divergence of parallel electron momentum term and the RHS. (b, d, f) A comparison of $dj_{\parallel}/dt$ and the RHS of Eq. \eqref{eq:current_eq_full_final_appendix}. The time traces are taken in the vicinity of the lowest order rational surface where (a, b, c) correspond to $x=-1.2 \rho_{i}$, (d, e, f) correspond to $x=0 \rho_{i}$, (g, h, i) correspond to $x=1.2 \rho_{i}$. The horizontal dashed lines in (e) and (f) indicate time trace averages.}
\label{fig:QFlat_s01_djdtRHS_testCase}
\end{figure}

Figure \ref{fig:QFlat_s01_djdtRHS_testCase} illustrates the agreement between LHS and RHS of Eq. \eqref{eq:current_eq_full_final_appendix} in a gyrokinetic simulation with strongly driven ITG turbulence. 
While the match is good, the small mismatch in subfigures (c,f,i) results from the finite resolution in $z$ and $x$ of the simulation. 
In this example, it is clear that term \textbf{(1)} in Eq. \eqref{eq:current_eq_full_final_appendix} dominates the RHS while terms \textbf{(3)} and \textbf{(4)} cancel out and terms \textbf{(2)} and \textbf{(5)} are subdominant. 
Therefore, we conclude that the main mechanism driving parallel current is the divergence of the parallel electron momentum flux.

\section{Modification of the safety factor profile by the turbulent $A_{\parallel}$}\label{app:turb_mod_q}
Here we present a derivation for how the imposed safety factor profile is modified due to turbulence-generated parallel currents.
In the following derivation, we are interested in modifications to the background safety factor profile.
Thus we focus solely on effects resulting from the time-averaged zonal component of $A_{\parallel}$.
This ensures that modifications to the magnetic field are confined to adjustments in the field line pitch on the flux surfaces, without altering the flux surface's overall shape.

Thus, the perturbed magnetic field can be expressed in terms of the parallel component $A_{\parallel}$ of the vector potential $ \mathbf{A} = A_{\parallel} \hat{\mathbf{b}}$, where $\hat{\mathbf{b}}$
\begin{equation}
\label{eq:B_perp_through_A_perp}
    \delta \mathbf{B}_{\perp} = \nabla \times \langle \langle \mathbf{A}_{\parallel} \rangle_{y} \rangle_{t}= \frac{\partial \langle A_{\parallel} \rangle_{y,t}}{ \partial x} \nabla x \times \hat{\mathbf{b}},
\end{equation}
where $\langle ... \rangle_{y}$ is the average over $y$ to select the zonal component and $\langle ... \rangle_{t}$ is the time average.

We begin by expressing the magnetic field 
\begin{equation}
\label{eq:B_total}
    \mathbf{B}_{total} = \mathbf{B}_{0}+\delta \mathbf{B}_{\perp}
\end{equation}
as a sum of the imposed background field $\mathbf{B}_{0}$ and a perturbed field $\delta \mathbf{B}_{\perp}$ arising from turbulence.
We write the background field in Clebsh form as in \eqref{eq:magnetic_field}
\begin{equation}
\label{eq:Clebsh_coord_appendix}
    \mathbf{B}_{0} = C_{xy} \nabla x \times \nabla y.
\end{equation}
The total safety factor $q_{total}$ can be expressed in a general form as
\begin{equation}
\label{eq:q_total_general_expression}
    q_{total} = \frac{1}{2 \pi} \int_{0}^{2\pi} \frac{\mathbf{B}_{total} \cdot \nabla \zeta}{\mathbf{B}_{total} \cdot \nabla z} dz.
\end{equation}
Using Eq. \eqref{eq:B_total} the magnetic field in the integrand can be split as 
\begin{equation}
\label{eq:q_tot_fullform}
    \frac{\mathbf{B}_{total} \cdot \nabla \zeta}{\mathbf{B}_{total} \cdot \nabla z} = \frac{(\mathbf{B}_{0}+\delta \mathbf{B}_{\perp})\cdot \nabla \zeta}{(\mathbf{B}_{0}+\delta \mathbf{B}_{\perp}) \cdot \nabla z} = \frac{\mathbf{B}_{0}\cdot \nabla \zeta}{\mathbf{B}_{0} \cdot \nabla z} \left[ \frac{1+\frac{\delta \mathbf{B}_{\perp} \cdot \nabla \zeta}{\mathbf{B}_{0} \cdot \nabla \zeta}}{1+\frac{\delta \mathbf{B}_{\perp} \cdot \nabla z}{\mathbf{B}_{0} \cdot \nabla z}} \right].
\end{equation}
Consistent with the gyrokinetic ordering, we assume that the perturbations to the background magnetic field are small compared to the background field.

We can therefore Taylor expand Eq. \eqref{eq:q_tot_fullform} to first order and find
\begin{equation}
\label{eq:q_tot_step2}
    \frac{\mathbf{B}_{total} \cdot \nabla \zeta}{\mathbf{B}_{total} \cdot \nabla z} =  q_{0} + \frac{\delta \mathbf{B}_{\perp} \cdot (\nabla \zeta - q_{0} \nabla z)}{\mathbf{B} \cdot \nabla z}.
\end{equation}

To arrive at an expression in the flux tube coordinates, we express $\nabla \zeta$, following the definition of the binormal coordinate in Eq. \eqref{eq:coordinates_flux_tube}, as
\begin{equation}
    \nabla \zeta = - \frac{1}{C_{y}} \nabla y + z \frac{d q_{0}}{ d x} \nabla x + q_{0} \nabla z.
\end{equation}
Inserting this expression together with Eqs. \eqref{eq:B_perp_through_A_perp} and  \eqref{eq:q_tot_step2} into Eq. \eqref{eq:q_total_general_expression} we arrive at
\begin{equation}
\label{eq:qtotal_step3}
    q_{total} = q_{0} - \frac{1}{2 \pi} \int_{0}^{2\pi} \frac{\partial \langle A_{\parallel} \rangle_{y,t}}{ \partial x} \frac{(\nabla x \times \hat{\mathbf{b}})\cdot \nabla y }{C_{y} \mathbf{B} \cdot \nabla z} dz.
\end{equation}
From Eq. \eqref{eq:Clebsh_coord_appendix} and $\mathbf{B} = B \hat{\mathbf{b}}$ we see that
\begin{equation}
    (\nabla x \times \hat{\mathbf{b}}) \cdot \nabla y = - \frac{B}{C_{xy}}.
\end{equation}
The unit vector in the magnetic field direction can be expressed as  \cite{Gorler2009Thesis}
\begin{equation}
    \hat{\mathbf{b}} = \frac{\mathbf{B}}{B} = \frac{1}{\sqrt{\gamma_{1}}} \nabla x \times \nabla y =  \frac{1}{J_{xyz} \sqrt{\gamma_{1}}} \mathbf{e}_{z}.
\end{equation}
Substituting the above two expressions and $\mathbf{e}_{z} \cdot \nabla z = 1$ in Eq. \eqref{eq:qtotal_step3} we finally arrive at
\begin{equation}
\label{eq:qtotal_final_app}
    q_{total} = q_{0} + \frac{1}{2 \pi} \int_{0}^{2\pi} \frac{\partial \langle A_{\parallel} \rangle_{y,t}}{ \partial x}  \frac{J_{xyz} \sqrt{\gamma_{1}}}{C_{xy} C_{y} } dz.
\end{equation}
Writing out the flux surface and time averages explicitly we see that a steady zonal $A_{\parallel}$ will modify the $q$ profile by adding the term
\begin{equation}
    \Tilde{q}_{A_{\parallel}} (x) = \frac{1}{2 \pi} \int_{0}^{2\pi} \frac{\partial \langle A_{\parallel} \rangle_{y,t}}{ \partial x} \frac{J_{xyz} \sqrt{\gamma_{1}}}{C_{xy} C_{y} } dz,
\end{equation}
appearing in Eq. \eqref{eq:total_mod_q}

\section{Simulation parameters}\label{app:simulation_parameters}
In this appendix, we present the parameters of our numerical simulations using GENE v2. All simulations use Miller local equilibrium geometry \cite{Miller1998}. In the tables below, $\epsilon$ is the inverse aspect ratio, $T_{i}/T_{e}$ is the ion to electron temperature ratio, $m_{i}/m_{e}$ is the ion to electron mass ratio, $L_{v}$ and $L_{w}$ are the extensions of the simulation domain in $v_{\parallel}$ and $\mu$ respectively. Lastly, $N_{x}$, $N_{y}$, $N_{z}$, $N_{v_{\parallel}}$, $N_{\mu}$ are number of grid points in the $x$, $y$, $z$, $v_{\parallel}$ and $\mu$ directions respectively, and $N_{species}$ is the number of species. Finally, $M \in \mathbb{N}$ determines the domain size according to the quantization condition of Eq. \eqref{eq:domain_quantization}.


\begin{table}[hb!]
\caption{Key parameters for the nonlinear plasma $\beta$ scan simulations with linear safety factor profile. The corresponding results are shown in Figs. \ref{fig:QFlat_s01_Apar_betaScan_070524}, \ref{fig:QFlat_s01_biCorr_b0vsb0001_paper_160524}, \ref{fig:QFlat_s01_Qtrace_betaScan_080524} and \ref{fig:QFlat_s01_b0vsb0001_ProfileCorrugations_300524}.}
\label{tab:parameters_nonlinear_linQ_s01_highres}
\begin{tabular}{|llll|}
\hline
\multicolumn{1}{|l|}{$\epsilon = 0.18$} & \multicolumn{1}{l|}{$q_{0}=1.4$} & \multicolumn{1}{l|}{$\hat{s} = 0.1$} & \multicolumn{1}{l|}{$\beta \in [0, 10^{-3}]$} \\ \hline 
\multicolumn{1}{|l|}{$m_{i}/m_{e}=3670$} & \multicolumn{1}{|l|}{$T_{i}/T_{e}=1$} & \multicolumn{1}{l|}{$R/L_{N} = 2.22 $} & \multicolumn{1}{l|}{$R/L_{T_{i}}=6.96$} \\ \hline 
\multicolumn{1}{|l|}{$R/L_{T_{e}} = 6.96$} & \multicolumn{1}{|l|}{$N_{pol}=1$} & \multicolumn{1}{|l|}{$L_{z}=2\pi$} & \multicolumn{1}{l|}{$L_{v}$=3 $(2T_{s}/m_{s})^{1/2}$} \\ \hline 
\multicolumn{1}{|l|}{$L_{w}$=12 $T_{s}/B_{ref}$} & \multicolumn{1}{l|}{$k_{y,min} \rho_{i} = 0.05$} & \multicolumn{1}{l|}{$L_{x} = 196.733 \rho_{i}$} & \multicolumn{1}{l|}{$\Delta y_{0} = 0$} \\ \hline
\multicolumn{4}{|l|}{$N_{x} \times N_{y} \times N_{z} \times N_{v_{\parallel}} \times N_{\mu} \times N_{species} = 384 \times 96 \times 16 \times 64 \times 9 \times 2$.} \\ \hline
\end{tabular}
\end{table}


\begin{table}[hb!]
\caption{Key parameters for the nonlinear study where the linear safety factor profile was adjusted by eliminating zonal $A_{\parallel}$ or imposing non-uniform safety factor profile perturbation. The corresponding results are shown in Figs. \ref{fig:QFlat_s01_mex50_Qtrace_betaScan_080524} and \ref{fig:QFlat_s01_mex50_Npol5_CorrComp}. A similar resolution was used for the simulations in Fig. \ref{fig:QFlat_s01_AE_linearVSsteppedQ_051424}, except $\beta = 0$ and electrons were treated adiabatically.}
\label{tab:parameters_nonlinear_linQ_s01_lowres_mex20}
\begin{tabular}{|llll|}
\hline
\multicolumn{1}{|l|}{$\epsilon = 0.18$} & \multicolumn{1}{l|}{$q_{0}=1.4$} & \multicolumn{1}{l|}{$\hat{s} = 0.1$} & \multicolumn{1}{l|}{$\beta = 0$ or $10^{-3}$} \\ \hline 
\multicolumn{1}{|l|}{$m_{i}/m_{e}=184$} & \multicolumn{1}{|l|}{$T_{i}/T_{e}=1$} & \multicolumn{1}{l|}{$R/L_{N} = 2.22 $} & \multicolumn{1}{l|}{$R/L_{T_{i}}=6.96$} \\ \hline 
\multicolumn{1}{|l|}{$R/L_{T_{e}} = 6.96$} & \multicolumn{1}{|l|}{$N_{pol}=1$ or $5$} & \multicolumn{1}{|l|}{$L_{z}=2\pi N_{pol}$} & \multicolumn{1}{l|}{$L_{v}$=3 $(2T_{s}/m_{s})^{1/2}$} \\ \hline 
\multicolumn{1}{|l|}{$L_{w}$=12 $T_{s}/B_{ref}$} & \multicolumn{1}{l|}{$k_{y,min} \rho_{i} = 0.05/N_{pol}$} & \multicolumn{1}{l|}{$L_{x} = 196.733 \rho_{i}$} & \multicolumn{1}{l|}{$\Delta y_{0} = 0$} \\ \hline
\multicolumn{4}{|l|}{$\Tilde{s}^{C} = (0.002, -0.0084, 0.0066, -0.1176, -0.0094, 0.0078)$, $\Tilde{s}^{C}_{n} = 0$ $\forall n>6$ } \\ \hline
\multicolumn{4}{|l|}{$\Tilde{s}^{S} = (0, 0, 0, 0.0057)$, $\Tilde{s}^{S}_{n} = 0$ $\forall n>4$} \\ \hline
\multicolumn{4}{|l|}{$N_{x} \times N_{y} \times N_{z} \times N_{v_{\parallel}} \times N_{\mu} \times N_{species} = 256 \times 48N_{pol} \times 16N_{pol} \times 32 \times 9 \times 2$.} \\ \hline
\end{tabular}
\end{table}


\begin{table}[hb!]
\caption{Key parameters for the nonlinear simulations with a small safety factor profile non-uniformity. The corresponding results are shown in Figs. \ref{fig:QFlat_s01_Apar_betaScan_070524}, \ref{fig:QFlat_s01_biCorr_b0vsb0001_paper_160524}, \ref{fig:QFlat_s01_Qtrace_betaScan_080524} and \ref{fig:QFlat_s01_b0vsb0001_ProfileCorrugations_300524}.}
\label{tab:parameters_nonlinear_sSsmall_study}
\begin{tabular}{|llll|}
\hline
\multicolumn{1}{|l|}{$\epsilon = 0.18$} & \multicolumn{1}{l|}{$q_{0}=2$} & \multicolumn{1}{l|}{$\hat{s} = 0$} & \multicolumn{1}{l|}{$\beta = 0$ or $10^{-3}$} \\ \hline 
\multicolumn{1}{|l|}{$m_{i}/m_{e}=368$} & \multicolumn{1}{|l|}{$T_{i}/T_{e}=1$} & \multicolumn{1}{l|}{$R/L_{N} \in \{ 0, 2.22 \} $} & \multicolumn{1}{l|}{$R/L_{T_{i}}=6.96$} \\ \hline 
\multicolumn{1}{|l|}{$R/L_{T_{e}} \in \{0, 6.96\}$} & \multicolumn{1}{|l|}{$N_{pol}=1$} & \multicolumn{1}{|l|}{$L_{z}=2\pi$} & \multicolumn{1}{l|}{$L_{v}$=3 $(2T_{s}/m_{s})^{1/2}$} \\ \hline 
\multicolumn{1}{|l|}{$L_{w}$=12 $T_{s}/B_{ref}$} & \multicolumn{1}{l|}{$k_{y,min} \rho_{i} = 0.05$} & \multicolumn{1}{l|}{$L_{x} = 150 \rho_{i}$} & \multicolumn{1}{l|}{$\Delta y_{0} = 0$} \\ \hline
\multicolumn{4}{|l|}{$\Tilde{s}^{C}_{n} = 0$ $\forall n$} \\ \hline
\multicolumn{4}{|l|}{$\Tilde{s}^{S}_{1} \in \{-0.024, -0.012, -0.006, -0.004, -0.002, 0 \}$, $\Tilde{s}^{S}_{n} = 0$ $\forall n>1$} \\ \hline
\multicolumn{4}{|l|}{$N_{x} \times N_{y} \times N_{z} \times N_{v_{\parallel}} \times N_{\mu} \times N_{species} = 192 \times 64 \times 24 \times 64 \times 12 \times 2$.} \\ \hline
\end{tabular}
\end{table}


\begin{table}[hb!]
\caption{Key parameters for the nonlinear simulations with a non-uniform safety factor profile and a scan in $\Delta y_{0}$. The corresponding results are shown in Figs. \ref{fig:QFlat_sSm0025_etaScan_qProfile_v2} and \ref{fig:QFlat_sSm0025_s0_Qtrace_etaScan}.}
\label{tab:parameters_nonlinear_sS_etaScan_study}
\begin{tabular}{|llll|}
\hline
\multicolumn{1}{|l|}{$\epsilon = 0.18$} & \multicolumn{1}{l|}{$q_{0}=2$} & \multicolumn{1}{l|}{$\hat{s} = 0$} & \multicolumn{1}{l|}{$\beta = 10^{-4}$} \\ \hline 
\multicolumn{1}{|l|}{$m_{i}/m_{e}=368$} & \multicolumn{1}{|l|}{$T_{i}/T_{e}=1$} & \multicolumn{1}{l|}{$R/L_{N} = 2.22 $} & \multicolumn{1}{l|}{$R/L_{T_{i}}=6.96$} \\ \hline 
\multicolumn{1}{|l|}{$R/L_{T_{e}} = 6.96$} & \multicolumn{1}{|l|}{$N_{pol}=1$} & \multicolumn{1}{|l|}{$L_{z}=2\pi$} & \multicolumn{1}{l|}{$L_{v}$=3 $(2T_{s}/m_{s})^{1/2}$} \\ \hline 
\multicolumn{1}{|l|}{$L_{w}$=12 $T_{s}/B_{ref}$} & \multicolumn{1}{l|}{$k_{y,min} \rho_{i} = 0.05$} & \multicolumn{1}{l|}{$L_{x} = 300 \rho_{i}$} & \multicolumn{1}{l|}{$\Delta y_{0} \in [2.5 \rho_{i}, 15 \rho_{i}]$} \\ \hline
\multicolumn{4}{|l|}{$\Tilde{s}^{C}_{n} = 0$ $\forall n$} \\ \hline
\multicolumn{4}{|l|}{$\Tilde{s}^{S}_{1} = -0.025$, $\Tilde{s}^{S}_{n} = 0$ $\forall n>1$} \\ \hline
\multicolumn{4}{|l|}{$N_{x} \times N_{y} \times N_{z} \times N_{v_{\parallel}} \times N_{\mu} \times N_{species} = 384 \times 64 \times 32 \times 64 \times 12 \times 2$.} \\ \hline
\end{tabular}
\end{table}


\begin{table}[hb!]
\caption{Key parameters for the nonlinear simulations with a non-uniform safety factor profile and a scan in $\Delta y_{0}$. The corresponding results are shown in Figs. \ref{fig:Qtrace_CBC_b0vsb0001_s01_sSm014_eta28_120824}, \ref{fig:QFlat_sSm014_etaScan_plasmaProfile_nomVS70pGrad_v2_241024} and \ref{fig:QFlat_CBC_b0vsb0001_s01_sSm014_eta028_plasmaProfile_120824}.}
\label{tab:parameters_nonlinear_pseudoglobal_study}
\begin{tabular}{|llll|}
\hline
\multicolumn{1}{|l|}{$\epsilon = 0.18$} & \multicolumn{1}{l|}{$q_{0}=2$} & \multicolumn{1}{l|}{$\hat{s}_{0} = 0.1$} & \multicolumn{1}{l|}{$\beta = 10^{-3}$} \\ \hline 
\multicolumn{1}{|l|}{$m_{i}/m_{e}=368$} & \multicolumn{1}{|l|}{$T_{i}/T_{e}=1$} & \multicolumn{1}{l|}{$R/L_{N} \in \{1.554, 2.22\}$} & \multicolumn{1}{l|}{$R/L_{T_{i}} \in \{4.872, 6.96\}$} \\ \hline 
\multicolumn{1}{|l|}{$R/L_{T_{e}} \in \{4.872, 6.96\}$} & \multicolumn{1}{|l|}{$N_{pol}=1$} & \multicolumn{1}{|l|}{$L_{z}=2\pi$} & \multicolumn{1}{l|}{$L_{v}$=3 $(2T_{s}/m_{s})^{1/2}$} \\ \hline 
\multicolumn{1}{|l|}{$L_{w}$=12 $T_{s}/B_{ref}$} & \multicolumn{1}{l|}{$k_{y,min} \rho_{i} = 0.05$} & \multicolumn{1}{l|}{$L_{x} = 196.7 \rho_{i}$} & \multicolumn{1}{l|}{$\Delta y_{0} \in [2.5 \rho_{i}, 15 \rho_{i}]$} \\ \hline
\multicolumn{4}{|l|}{$\Tilde{s}^{C}_{n} = 0$ $\forall n$} \\ \hline
\multicolumn{4}{|l|}{$\Tilde{s}^{S}_{1} = -0.14$, $\Tilde{s}^{S}_{n} = 0$ $\forall n>1$} \\ \hline
\multicolumn{4}{|l|}{$N_{x} \times N_{y} \times N_{z} \times N_{v_{\parallel}} \times N_{\mu} \times N_{species} = 288 \times 64 \times 24 \times 64 \times 12 \times 2$.} \\ \hline
\end{tabular}
\end{table}

\clearpage
\section*{References}

\bibliography{nfref.bib}

\end{document}